\documentclass[compsoc,conference,a4paper,10pt,times]{IEEEtran}
\IEEEoverridecommandlockouts
\usepackage{cite}
\usepackage{amsmath,amssymb,amsfonts}
\usepackage{algorithmic}
\usepackage{graphicx}
\usepackage{textcomp}
\usepackage{bmpsize}
\usepackage{xcolor}
\usepackage{lipsum}
\usepackage{xspace}
\usepackage{subfigure}
\usepackage{enumitem}

\usepackage{url}
\usepackage{breakurl}

\usepackage[colorlinks=true,urlcolor=black]{hyperref}
\def\BibTeX{{\rm B\kern-.05em{\sc i\kern-.025em b}\kern-.08em
    T\kern-.1667em\lower.7ex\hbox{E}\kern-.125emX}}
\begin{document}

\newcommand{\name}{ephemeral\xspace}
\newcommand{\Name}{Ephemeral\xspace}
\newcommand{\Secref}[1]{\S\ref{#1}}
\newcommand{\Figref}[1]{Fig.~\ref{#1}}
\newcommand{\descr}[1]{{\bigskip\noindent\textbf{#1}}}
\newcommand{\descrplain}[1]{{\smallskip\noindent\textbf{#1}}}

\title{Ephemeral Astroturfing Attacks: The Case of Fake Twitter Trends\\ {\footnotesize
    } \thanks{NB: appendices, if any, did
    not benefit from peer review.}}

\author{\IEEEauthorblockN{Tuğrulcan Elmas}
\IEEEauthorblockA{\textit{EPFL} \\
Lausanne, Switzerland \\
tugrulcan.elmas@epfl.ch}
\and
\IEEEauthorblockN{Rebekah Overdorf}
\IEEEauthorblockA{\textit{EPFL} \\
Lausanne, Switzerland \\
rebekah.overdorf@epfl.ch}
\and
\IEEEauthorblockN{Ahmed Furkan Özkalay}
\IEEEauthorblockA{\textit{EPFL} \\
Lausanne, Switzerland \\
ahmed.ozkalay@epfl.ch}
\and
\IEEEauthorblockN{Karl Aberer }
\IEEEauthorblockA{\textit{EPFL} \\
Lausanne, Switzerland \\
karl.aberer@epfl.ch}
}

\maketitle

\begin{abstract}
We uncover a previously unknown, ongoing astroturfing attack on the popularity mechanisms of social media platforms: \name astroturfing attacks. In this attack, a chosen keyword or topic is artificially promoted by coordinated and inauthentic activity to appear popular, and, crucially, this activity is removed as part of the attack. We observe such attacks on Twitter trends and find that these attacks are not only successful but also pervasive. We detected over 19,000 unique fake trends promoted by over 108,000 accounts, including not only fake but also compromised accounts, many of which remained active and continued participating in the attacks. Trends astroturfed by these attacks account for at least 20\% of the top 10 global trends. Ephemeral astroturfing threatens the integrity of popularity mechanisms on social media platforms and by extension the integrity of the platforms. 
\end{abstract}
\section{Introduction}

Mechanisms deployed by social media platforms to display popular content are a primary vector by which platforms increase engagement. Facebook's newsfeed algorithm; Reddit's ``r/popular''; and  Twitter's trending topics, ``trends,'' are integral to both platform functionality and the underlying business model. These mechanisms are valuable because they determine which content is most visible to users. Twitter's \emph{trends} can be equated to traditional advertising channels and can be useful for marketing~\cite{carrascosa2013trending}, as Twitter acknowledges by charging companies to promote their brands on trends for a day~\cite{trendcost}.

The integrity of such popularity mechanisms is integral to the social media ecosystem. Users expect that the popular content they are shown is the result of authentic activity on the platform, legitimate grassroots campaigns expect that their content will be fairly considered, and the platform expects that showing popular content increases engagement. Further downstream, advertisers expect that popularity mechanisms behave in a way to increase engagement and therefore revenue. Even further, those who use trends to study society and social media, i.e. researchers and journalists, expect that trends accurately reflect popular themes that are discussed by the public.  

Since these popularity mechanisms carry so much influence and potential for revenue, they are an attractive target for adversaries who want their illicit content to be seen by many users. For instance, ``like farms'' are used to generate fake likes on Facebook to boost posts to the top of users' news feeds~\cite{Cristofaro}, and bots can be used on Reddit to artificially ``upvote'' posts to increase their visibility~\cite{carman2018manipulating}. In the case of Twitter trends, adversaries, sometimes from abroad~\cite{caucuses}, boost disinformation and conspiracy theories to make them trend so that they are further amplified~\cite{epstein}, as in the case of QAnon followers hijacking the trend \#SaveTheChildren~\cite{qanon}. Due to this incident, many called on Twitter to stop curating trends by using the hashtag \#UntrendOctober~\cite{untrendoctober}. 

Attacks on popularity mechanisms rely on making inauthentic content or actions appear organic. Borrowing terminology used to refer to campaigns that fake grassroots organizing on social media, we call them ``astroturfing'' attacks. Once exposed, astroturfing attacks erode user trust in the platform. Gaining an understanding of these attacks is a crucial part of keeping the platforms safe for users and valuable for advertisers, thus preserving the business model of the platforms. 

In this paper, we provide an in-depth analysis of a new type of astroturfing attack that remains unstudied in the academic literature which we call \emph{\name astroturfing}. \Name astroturfing differs from traditional astroturfing in that the actors hide the malicious activity while successfully executing an astroturfing attack, paradoxically aiming to make something more visible while making the content responsible for the visibility invisible. By removing any evidence of the attack, ephemeral astroturfing outperforms other approaches in three key ways: (i) it enables the use of active, compromised accounts as sources of fake interactions, accelerating the popularity; (ii) it evades detection by users, the platform, and academic studies; and (iii) it prevents users from reporting the malicious activity as spam, so traditional spam classifiers are unable to prevent future attacks. 

We focus on fake Twitter trends as a case study to investigate \name astroturfing attacks. Twitter is a popular platform for many critical discussions, including political debates, with appropriate data available to study: Twitter provides both deletion notices and trends through its official APIs. We observe that Twitter trends suffer from \name astroturfing attacks both in Turkish local trends, affecting Turkey's 11.8 million active users, and global trends. Precisely, we find that ephemeral astroturfed attacks on Twitter trends started in 2015 and accounted for at least 47\% of the top-5 daily trends in Turkey and at least 20\% of the top 10 global trends. We find that Twitter does not consider whether a tweet has been deleted when determining which keywords should trend and thus is vulnerable to \name attacks.

Ephemeral astroturfing is enabled by the current design of the algorithm that determines Twitter trends. Trends are refreshed every 5 minutes, taking as input tweets that have been published in some time interval. However, despite the importance of the integrity of the list of trends, the algorithm \emph{does not check} whether those tweets are still available or have been deleted. This vulnerability can be expressed as a sort of Time-of-Check-Time-of-Use (TOCTOU) attack, by which at the moment that the data is ``used'' to determine a trend, it is different than when it was ``checked'' because it is deleted. In other words, this attack exploits a violation of the complete mediation principle when using security-critical inputs (tweets) to update a key asset for the platform. 

Due to the severity of the attack, we notified Twitter (once in July 2019 and again in June 2020) and provided a detailed description of the attack and the accounts involved. After the first notification they acknowledged that the attacks do exist (July 2019), and after the second notification (June 2020) they replied that they would forward them to the relevant team to address. We have followed up since, but have not received any indication that they are progressing. The attacks on Twitter trends continue as of February 2021.

In summary, our contributions are the following: 
\begin{itemize}[nosep]
    \item We define and describe a new type of attack on the popularity mechanisms: \emph{\name astroturfing}  (\Secref{sec:definition}).
  
    \item We uncover \name astroturfing on Twitter trends as it occurs in-the-wild. We find that it has been ongoing since 2015 and that it has a strong influence on local trends i.e., we find more than {19,000 unique keywords} that are the result of \name astroturfing attacks (\Secref{sec:turkish}) which employed at least {108,000 bots;} and on global trends, i.e., we find that at least 20\% of the popular global trends during our study were the result of \name astroturfing (\Secref{sec:attack_analysis}). Our study is the \emph{first large-scale analysis} of fake trends.
    
    \item We study the ecosystem behind \name astroturfing attacks on Twitter trends. We find that they rely on a mix of bots and compromised accounts (\Secref{sec:useranalysisshort}). We also find that there is a business model built around the attacks in (\Secref{sec:trendanalysis}).

    \item We discuss the implications on platform security and society, propose countermeasures, and identify barriers to deploying defenses in practice. (\Secref{sec:defenses}).
    
\end{itemize}

\section{Background and Related Work}
\label{sec:related}

\noindent\textbf{Social Media Manipulation}
The wide adoption of social media platforms has attracted adversaries aiming to manipulate users on a large scale for their own purposes. Such manipulation attacks span from targeted advertising assisted by mass data collection~\cite{cadwalladr2018revealed} to state-sponsored trolling~\cite{nyst2018state}, propaganda~\cite{thomas2012adapting}, spam~\cite{grier2010spam, spamcampaign, impacts2018, yardi2010detecting}, popularity inflation~\cite{fameforsale}, and hashtag hijacking~\cite{vandam2016detecting}. Many of these manipulation attacks employ bots and bot-nets to execute since wide deployment is often a necessary component. We focus on this class of bot-assisted manipulation attacks.  In our study, we observed political propaganda (not necessarily pro-government) and illicit advertisements that manifest themselves not through hashtag hijacking, as is often the case as well, but through direct trend manipulation.

Bots are becoming increasingly difficult to identify manually~\cite{infiltrating,turing} or automatically~\cite{shift,onthecapability}. Social bots are designed to mimic human users on social media~\cite{firstmonday}; they copy real identities (personal pictures, tweets), mimic the circadian rhythm of humans, gain followers by following each other, and mix malicious and hand-crafted tweets~\cite{evasion}. CyboHuman bots~\cite{areyoucyborg}, \emph{cyborgs}, humans assisted by bots~\cite{augmented}, and \emph{augmented humans} mix automation and human activity. In some cases, users register their accounts with malicious apps that make them part of a botnet. \Name astroturfing attacks allow attackers to employ compromised users who continue using the account in parallel with the attackers, similar to~\cite{businessinsider}. Attackers hide from the legitimate user by deleting the attack tweets. Since these are otherwise benign users, they are likely to confuse supervised methods due to their dissimilarities to traditional bots. They would also confuse graph-based detection systems such as ~\cite{yu2006sybilguard,jia2017random} since they connect with other benign accounts. Although they are compromised, compromised account detection systems~\cite{kuzenimyazmis,compa,towards,authorshipverification} are not able to detect them if they do not account for deletions since the tweets that disclose compromisation are deleted.

Existing bot detection methods fall short of detecting the bot behavior we describe here as they rarely consider content deletion. Botometer~\cite{varol1} works on a snapshot of a profile and not on real-time activity, so it cannot detect the bot-like activity of accounts analyzed in this study since such activity is deleted quickly. Recently, Varol et al. ~\cite{varol2} used content deletion as a bot feature but used a proxy to capture deletions: a high recent tweeting rate but a low number of tweets. This may capture the deletion of old tweets but not tweets deleted quickly. Debot~\cite{debot} is based on keyword filtering by Twitter's Streaming, which does not give deletion notices and would not collect the relevant data if the attacked keyword is not be provided before the attacks, which is not possible for the keywords which trend only once. Chavoshi et al.~\cite{farkeden} discovered a set of Turkish bots with correlated deletion activity to hide bot-like behavior. However, this study did not uncover whether these deletions were part of the astroturfing attacks we describe here. In our work, we classify the bot-created fake trends using their characteristic behavior:  deletions and the generated content.

\descrplain{Astroturfing and Fake Trends} Although astroturfing by attacking trends using bots and manufacturing fake trends has been briefly reported on by the news media in both Saudi Arabia~\cite{Bbctrending} and Turkey~\cite{hurriyet}, it remains understudied in the academic literature. To the best of our knowledge, this work is the first to systematically study the mechanics of manipulating the Twitter trends feature on a large scale. 

While not directly concerning the trending mechanism, previous works have analyzed \emph{campaigns} that are artificially promoted~\cite{koreantrolls,abuse, varoltrends,ferrara2017disinformation} or found evidence of manipulation of popular topics that are also trending by studying suspended and/or fake accounts and the overall temporal activity~\cite{manipulation, jones2019gulf}. They stopped short of studying malicious activity before keywords' reaching trends lists. In our work, we study the adversarial behavior that aims to push certain keywords to trends list directly, the behavior to evade the detection, and the accounts that are used for such operation.
\section{\Name Astroturfing}
\label{sec:definition}

\noindent\textbf{Basis for Definition} To define \name astroturfing, we look to the case of an attack that we observed in-the-wild that has been targeting Twitter trends in Turkey. To understand the attack, we created a honeypot account and signed it up for a free follower scheme that phishes users’ credentials. We suspected that this scheme was being used to compromise accounts for \name astroturfing attacks because the scheme was being advertised via \name astroturfing. Our suspicions were confirmed when our account began tweeting and quickly deleting content containing keywords of about-to-be trends.  Precisely, our astrobot account tweeted 563 times in 6 months before we exited the scheme. We now describe \name astroturfing attacks on Twitter based on our observations.

\descr{Attack Summary} The goal of this attack is clear: make a topic popular. On Twitter, this translates to a \emph{target keyword} reaching the \emph{trends lists}. In contrast, however, the methods hint at an attempt to remain stealthy.

\begin{figure*}[ht]
\centering     
\subfigure{\label{fig:a}\includegraphics[width=0.32\textwidth]{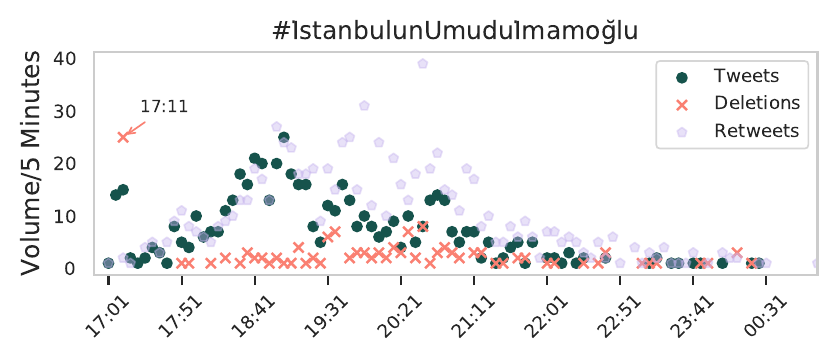}} 
\subfigure{\label{fig:c}\includegraphics[width=0.32\textwidth]{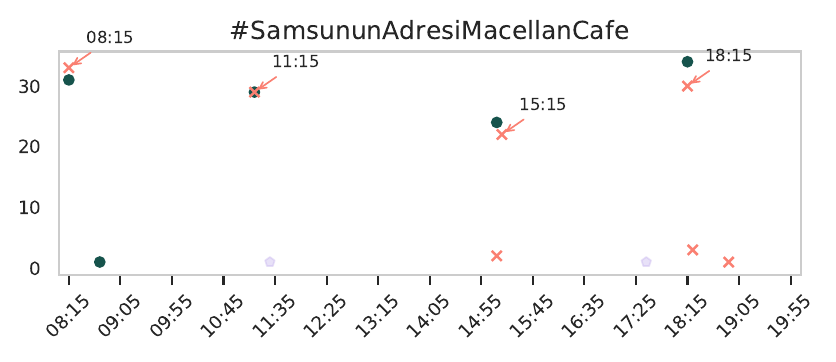}}
\subfigure{\label{fig:b}\includegraphics[width=0.32\textwidth]{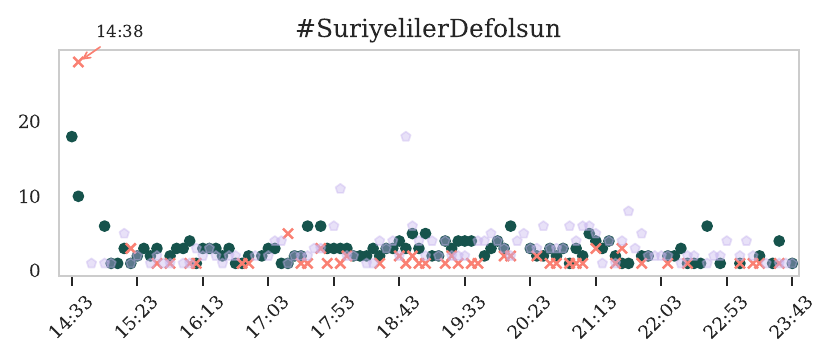}}

\caption{\small \emph{\#\.{I}stanbulunUmudu\.{I}mamo\u{g}lu} is a slogan associated with a candidate in the 2019 {I}stanbul election rerun. Note that although the hashtag is astroturfed by an attack initially (at 17:11), it was later adopted by popular users who got many retweets and drew the attention of the wider public. 
\emph{\#SamsununAdresiMacellanCafe} is an advertisement for a cafe, astroturfed to be seen in trends in Turkey. The hashtag did not receive attention from anyone other than astrobots: there are only coordinated tweets and deletions. 
\emph{\#SuriyelilerDefolsun} is a derogative slogan meaning "Syrians Get Out!". The hashtag grabbed the attention of the wider public due to its negative sentiment and sparked controversy in Turkey despite being astroturfed.
\label{fig:ex}}        
\end{figure*}

An \name astroturfing attack is executed by a number of accounts that are controlled by a single entity, which we refer to as \emph{astrobots}. Each astrobot creates a tweet at roughly the same time. After a short period, these tweets are deleted at roughly the same time. Alongside the target keyword, each tweet contains some pre-made or generated content that is enough to pass the spam filters of the platform (but not necessarily the Turing test). In the case of Twitter trends, we also observe that each account involved only posts with the target keyword once per attack. After an attack that renders a keyword trending successfully, other users adopt the new trend and post tweets that are not deleted. 
\Figref{fig:ex} shows the tweeting and deletion patterns of different astroturfing attacks with distinct non-adversarial behavior patterns.

\definecolor{green_dot}{HTML}{15534c}
\definecolor{yellow_x}{HTML}{a57d50}

\newcommand{\tweet}{t}
\newcommand{\tweetset}{T}
\newcommand{\content}{c}
\newcommand{\contentset}{C}
\newcommand{\software}{s}
\newcommand{\created}{p}
\newcommand{\createdset}{\mathcal{P}}
\newcommand{\deleted}{d}
\newcommand{\deletedset}{\mathcal{D}}
\newcommand{\attackmath}{\mathcal{A}}

\newcommand{\numtweetsthreshold}{\kappa}
\newcommand{\numgeotagged}{\beta}
\newcommand{\timeframethreshold}{\alpha_{p}}
\newcommand{\timeframethresholddel}{\alpha_{d}}
\newcommand{\lifetimethreshold }{\theta}

\descr{Attack Model} Let $w$ be the target keyword. Let a set of posts that contain $w$ be $\tweetset =\{\tweet_0, \tweet_1,...\tweet_n\}$, with creation time $\created_{\tweet_i} \in \createdset=\{\created_0, \created_1,...\created_n\}$, deletion time $\deleted_{\tweet_i}  \in \deletedset=\{\deleted_0, \deleted_1,...\deleted_n\}$. 
An attack $\attackmath$ occurs when there is a ${\tweetset}$ s.t.

\descrplain{1. Many posts}: $|\tweetset| > \numtweetsthreshold$:  at least $\numtweetsthreshold$ posts involved, 

\descrplain{2. Correlated Posting:} $max(\createdset) - min(\createdset) < \timeframethreshold$: the posts are created  within a window of size $\timeframethreshold$,

\descrplain{3. Inauthentic Content}: each post is comprised of $w$ and a premade or generated content $c$ that will pass the platform's spam filters, 
    
\descrplain{4. Correlated Deletions:} $max(\deletedset) - min(\deletedset) < \timeframethresholddel$: the posts are deleted  within a window of size $\timeframethresholddel$,
    
\descrplain{5. Quick Deletions:} $\deleted_{\tweet_i} - \created_{\tweet_i} < \lifetimethreshold \quad \forall{\tweet_i \in T}$: all posts are deleted within $\lifetimethreshold$.




We leave the parameters ($\numtweetsthreshold$,$\timeframethreshold$,$\timeframethresholddel$,$\lifetimethreshold$, $c$) 
in the definition unspecified and later infer concrete values based on the instances of the attack that we detect. 

To simulate trending behavior and confuse the algorithm which computes how popular the target keyword is, the attackers create many correlated posts in a short time window (rules 1 and 2).  Any type of coordinated and/or bot activity has to pass the spam filters to evade detection and also to be considered in the platforms' metrics for popularity. These attacks are too large and coordinated to be executed at scale with handcrafted content, so the content must be pre-made or generated by an algorithm and therefore exhibit patterns in their content (rule 3). While recent advances in generating meaningful text make it more difficult for humans to spot such patterns, these advances have not reached the point of being able to create short texts related to a keyword for which it has no training data. Additionally, such arrangements are costly. These three points are common to all astroturfing attacks. 

The ephemerality is captured by rules 4 and 5 in the attack model. Both appear to be the result of the attackers' tendency to quickly hide any trace of their attack from the public and the compromised accounts they employ. Additionally, deletions create a clean slate when users click on a trend, i.e., there will be no posts associated with the keyword when someone clicks on it on Twitter's trends list, so the attackers can post new content and be the first in the search results of the target keyword. 

\section{The Case of Fake Twitter Trends in Turkey}
\label{sec:turkish}
While astroturfing attacks are a global problem, we observe \name astroturfing on a large scale in local trends in Turkey. Turkey has the 5\textsuperscript{th} highest number of monthly Twitter users and a highly polarized political situation~\cite{turkishpolarization,mccoy2018polarization}. The Turkish mainstream media has occasionally reported about the prevalence of fake trends there~\cite{cengizsemerci,bbcturkce,hurriyet,habericin}, primarily sourced through interviews with attackers who manifest themselves as social media agencies. These agencies can be found via a simple Google Search for trend topic services and even advertise themselves using fake trends. 

We inspected the attack tweets used to create fake trends reported by Turkish media~\cite{yeniakit,medyafaresi,kaparoz}. Additionally, we inspected reports from Turkish Twitter users such as \textit{@BuKiminTTsi}, an account dedicated to reporting fake trends. We found a pattern in the structure of the deleted tweets: the content appears to be sourced from a lexicon of Turkish words and phrases, e.g. ``to organize milk frost deposit panel.'' They do not have a sentence structure nor do they convey meaning and the verbs are always in the infinitive form. We call these tweets \emph{lexicon tweets}. Our honeypot account also tweeted and deleted such tweets while promoting fake trends.

In this section, we uncover a massive \name astroturfing operation in Turkey. First, we inspect and annotate trends starting from 2013 and find the first instance of an \name astroturfing attack. Next, we show the features of the attack concerning our attack model. Finally, we build a training set and train a classifier to classify to find all trends that are associated with at least one attack.

\subsection{Datasets}
To study trends, we first need a trend dataset. We collect all trends in Turkey from an external provider\footnote{http://tt-history.appspot.com}. This list contains every trending keyword since July 7, 2013. The trends are collected every 10 minutes and indexed only by date, not time. As such, we treat every date and keyword pair as a separate trend to account for keywords trending in multiple days. 
Second, we need tweets. To this end, we employ Archive's Twitter Stream Grab\footnote{\url{https://archive.org/details/twitterstream} (accessed 2020-02-01)}, which contained 1\% of all Twitter data from September 2011 until September 2019 at the time of this analysis. This dataset contains deletions of tweets as well as the exact time the tweet is deleted. We verified that these deletions are due to authenticated users deleting tweets and not due to Twitter administrative actions by contacting Twitter. Our trend dataset does not contain the exact time a trend reaches trending but only the date. Therefore, for each trend, we associate tweets that contain the trend that is either posted on the same day that the keyword was trending or the day before to account for the keywords that were trending after midnight. (We later confirm that our results are robust to this design decision as most of the astroturfed trends do not have any previous discussions that stretch beyond a day earlier. See \Secref{sec:attack_analysis} for details.) We name this combined dataset the \emph{retrospective} dataset.

\subsection{Manual Annotation of Attacked Trends}
The goal of the manual annotation task is to uncover which keywords were trending as the result of an \name astroturfing attack and which were not. The annotators inspect trends, along with any tweets, deleted or otherwise, that contain the trending keyword. 

We first filter out trends with less than 10 associated tweets so that we are left with those that have enough data to meaningfully assign a label. Of those that remain, we randomly select one trend per day, resulting in 2,010 \emph{trend-date pairs} in total. 

The annotators examined the first 10 tweets of each trend and their deletion times, if available. A trend was considered to be initiated solely by an ephemeral astroturfing attack if 1) the tweets were deleted quickly and 2) the content of the first 10 associated tweets have a describable pattern that indicates automation (e.g. lexicon, random characters, repeated text). 

Note that constraining the annotation to only the first 10 tweets may hurt recall, i.e. we may miss the case where many tweets containing the target keyword are posted earlier in the same day of the attack so the attacked trend appears to be organic when only the first 10 tweets are considered. However, our observations and analyses in \Secref{sec:attack_analysis} show that this behavior is rare. 

Two authors contributed to the annotation process evenly. One author additionally annotated a random sample of 200 trends. The annotation agreement on whether a trend was initiated by an ephemeral astroturfing attack or not was k = 0.88 (almost perfect agreement). We further annotated the tweets associated with each of the 182 trends (5,701 tweets, 5,538 with unique content) as if they are part of an attack (i.e. if they are created and deleted shortly together while having the same pattern in their content) or not. Additionally, both annotators created subsets of the ``not \name astroturfing'' label for other types of astroturfing attacks (i.e. those which did not employ deletions). These attacks did not employ deletions so they are out of scope for this paper, but we include a brief supplementary analysis in Appx.~\ref{sec:historicity}.

We found that the first instance of a trend employing \name astroturfing attacks was in June 2015 and by 2017 it had become mainstream. Overall we found 182 trends that were astroturfed by \name astroturfing attacks using lexicon tweets. We did not observe any trends that are not promoted by lexicon tweets and still have the deletion patterns in our attack model.

\subsection{Analysis of Annotated Trends}

\begin{figure}[ht]
\begin{center}
\includegraphics[width=\columnwidth]{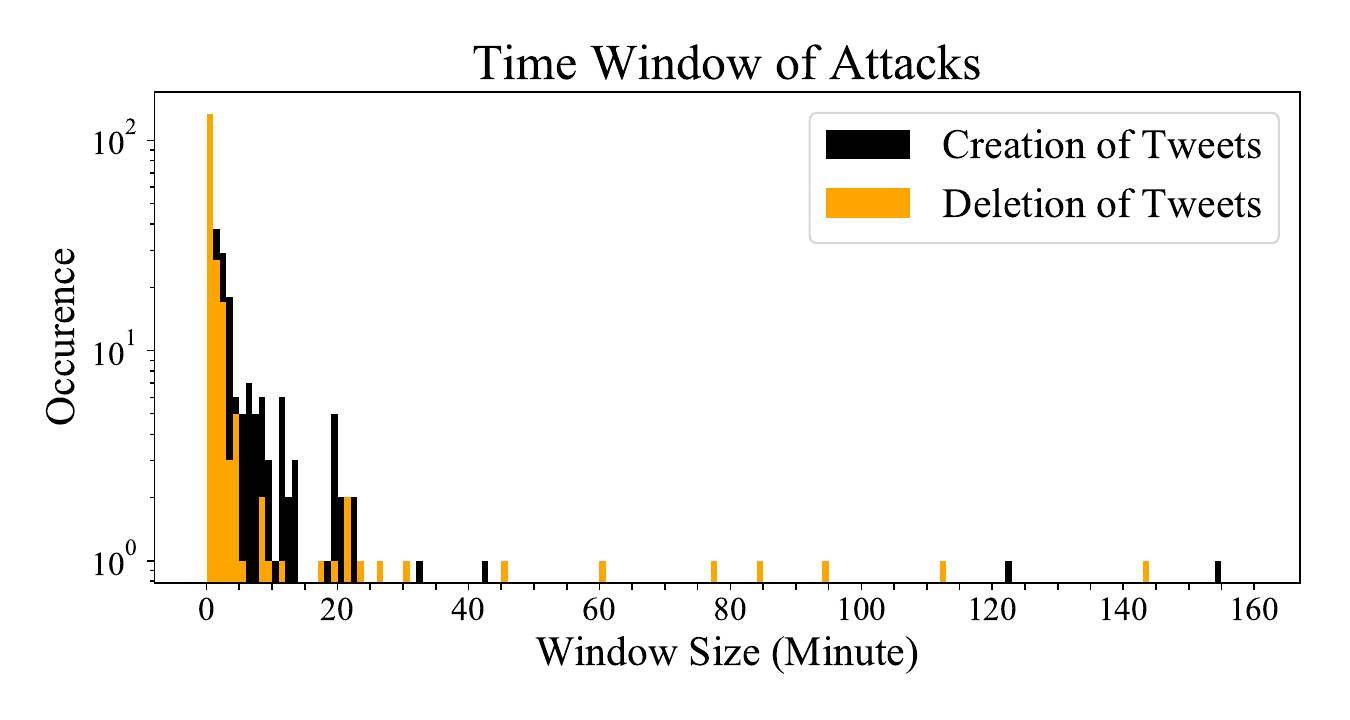}
\caption{The size of the time window in which the attack tweets are created ($<$ $\timeframethreshold$) is shown in blue. This shows the difference between the first and last tweet created containing the keyword for each trend. The size of the time window in which the attack tweets are deleted ($<$ $\timeframethresholddel$) is shown in orange. This shows the difference between the first and last tweet deleted containing the keyword for each trend. Most attacks occur in a very small time window. \label{fig:attackwindow_groundtruth_font20}}
\end{center}
\end{figure}

\noindent\textbf{Time Window of Actions}
Per our definition, \name astroturfing attacks post many attack tweets in a very small time window ($<$ $\timeframethreshold$) and delete them in a very small time window ($<$ $\timeframethresholddel$). \Figref{fig:attackwindow_groundtruth_font20} shows how small this time window is for the attacks we labeled: except for a few outliers, both $\timeframethreshold$ and $\timeframethresholddel$ are only a few minutes. 

\begin{figure}[htb]
\begin{center}
\includegraphics[width=\linewidth]{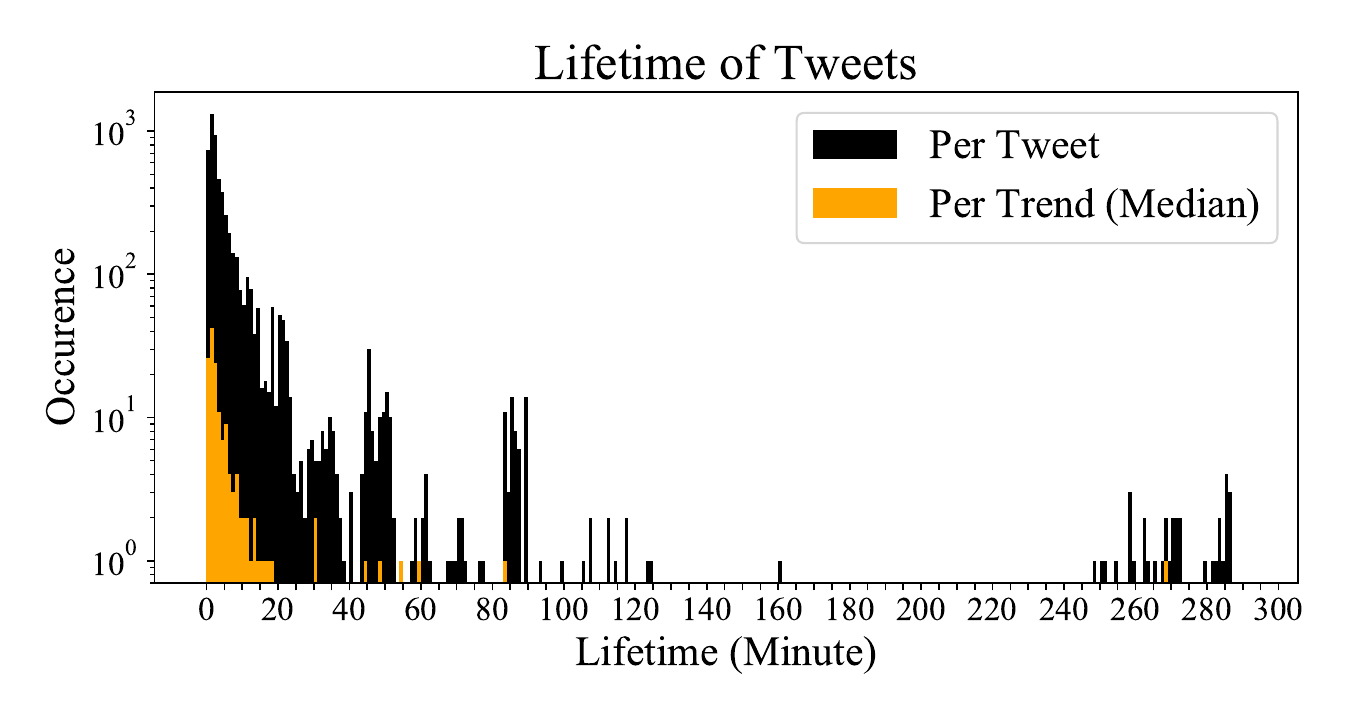}
\caption{Lifetime, the difference between time of creation and deletion of the annotated lexicon tweets. Blue shows the lifetime of individual tweets, and orange shows the median of the lifetime of tweets per trend. Attackers delete the tweets in 10 minutes (in most cases) and the difference between two histograms suggests that sometimes they miss some tweets to delete.}
\label{fig:lifetime_bar}
\end{center}
\end{figure}

\descrplain{Lifetime of Attacks}
\Name astroturfing attacks post many tweets and then delete them after a short period of time ($<$ $\lifetimethreshold$). \Figref{fig:lifetime_bar} shows the difference between the creation and deletion times of each attack tweet (i.e. lifetime, or how long a tweet ``lives'' before it is deleted) and the median lifetime of tweets per trend. Most have a very short median lifetime; however, some tweets are deleted after a few hours. This might be due to an error on the attackers' side (i.e. buggy code).

\subsection{Classification to Uncover More Attacks}

Next, we aim to automate the process of building a large-scale dataset of \name astroturfing attacks in order to perform a large-scale analysis. We build a simple classifier based on the features of the annotated data and the tweets collected from our honeypot account (\Secref{sec:definition}).

\descrplain{Lexicon Content}
\label{sec:incomprehensible}
Both our analysis of the annotated trends and our honeypot's tweets tell us that the \name astroturfing attacks that we see in this case employ lexicon tweets, which are trivial to classify. We study the honeypot's tweets to derive the rules for the lexicon classifier, since we are certain these tweets were sent by the attackers. We came up with the following rules and evaluated them on the 5,538 unique annotated lexicon tweets:

\begin{enumerate}[nosep]
\item Only alphabetical characters except parenthesis and emojis. (99.4\% of honeypot, 96.6\% of annotated). 
\item Beings with a lowercase letter. (99.4\% of honeypot, 96.3\% of annotated). False negatives were proper nouns from the lexicon.
    
\item Has between 2-9 tokens, excluding emojis. This range corresponds to the maximum and minimum number of tokens of the honeypot's tweets. In the annotation set, there were 5 lexicon tweets with only one token and 29 with more than 9. (100\% of honeypot, 99.4\% of annotated).
\end{enumerate}

The combination of these rules yields a recall of 92.9\% (5,147 / 5,538). To compute precision on deleted tweets, we ran the classifier on all of the deleted tweets in the sample of 2,010 trends: 17,437 tweets in total after dropping any duplicates (e.g. retweets). The classifier reported 370 lexicon tweets or a precision of 93.3\%. Of the false positives, 336 were from before June 2015, indicating that they were used in astroturfing attacks that predate the rise of \name astroturfing using lexicon tweets (see Appx.~\ref{sec:historicity} for details). There were only 34 false positives after June 2015.

\begin{figure}[htb]
\begin{center}
\includegraphics[width=\linewidth]{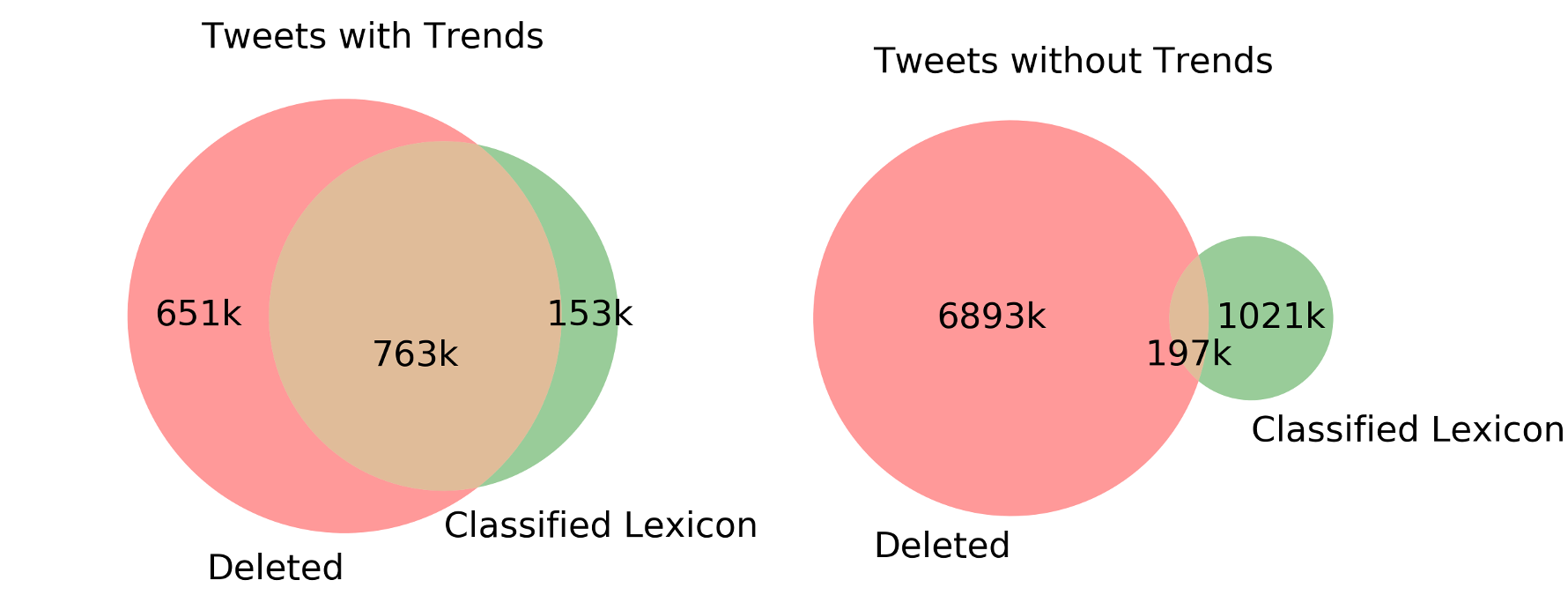}
\caption{Venn Diagram of the retrospective dataset concerning deleted and lexicon tweets. Tweets that are classified as lexicon account for only 2.3\% of all deleted tweets that are not associated with any trend (right diagram), but 53.1\% of all tweets associated with a trend (left diagram). Further, 83.2\% of all tweets that are classified as lexicon and associated with a trend are deleted. 
}
\label{fig:lexicon_stats}
\end{center}
\end{figure}

To corroborate the precision at scale, we show that lexicon tweets are common among deleted tweets associated with trends but rare otherwise. We classify all Turkish tweets in our retrospective dataset from June 2015. \Figref{fig:lexicon_stats} shows that most lexicon tweets associated with a trend are deleted, but very few lexicon tweets not associated with a trend are deleted.

Although lexicon tweets appear to be generated randomly using a lexicon of tweets, some occur more than once. Table \ref{tab:lexicon_examples} shows the most commonly repeated tweets (excluding the target keyword), their translations, and the number of times they occur in the data. We also observe that some words are so uncommon that even a native Turkish speaker may need to refer to a dictionary. This suggests that the attackers may be using infrequent words to pass Twitter's spam filters.

\begin{table*}[htb]
\caption{The most frequent lexicon tweets found in the dataset.}
\label{tab:lexicon_examples}
\centering
\resizebox{\textwidth}{!}{%
\begin{tabular}{cll}
\textbf{Frequency}  & \textbf{Tweet}                                                 & \textbf{Translation}                                                        \\ \hline
77 & tenkidi kayna\c{s}t{\i}rabilme siperisaika                   & critical to be able to boil lightning rod                                  \\ 
64 & yar{\i}m g\"{u}n g\"{u}zelle\c{s}tirilme oyalayabilme                & half day to be prettifiable to be able to distract                         \\ 
64 & kargocu yan bak{\i}\c{s} az{\i}msanma aforozlanma               & deliveryman side view to be underestimated to be excommunicated           \\ 
64 & yemenici kalsiyum klor\"{u}r yar{\i}m bağlaş{\i}m koyulaşt{\i}rmak & hand-printed head scarve maker chloride half coupling to coagulate         \\ 
62 & \"{o}rg\"{u}tleme s\"{u}t karlanmak panel                         & to organize milk frost deposit panel                                       \\ 
\end{tabular}%
}
\end{table*}

\descrplain{Supplementary Annotations}
Our annotated dataset contains only 182 astroturfed trends, which is too few to train a classifier. As such, we perform a second phase of annotations to extend the dataset. We selected a random sample of 5 trends per day after Jan 1, 2017, (4,255 trends in total) and annotate whether or not they were part of a lexicon attack. As this task is much larger than the previous one, and because we now have more information about how these attacks operate, to speed up annotations we only considered deleted tweets associated with a trend. We look for a burst of lexicon tweets posted and then deleted. We found that the condition \emph{at least four lexicon tweets} successfully differentiated attacked and organic trends, with only one instance of an attacked trend with fewer (3) lexicon tweets. Two annotators confirmed and corrected the labels. The resulting dataset contains 838 trends that were associated with at least one attack and 3,417 which were not which indicates a base rate of 19.7\% for the attacks.

\begin{figure}[ht]
    \centering
    \includegraphics[width=\columnwidth]{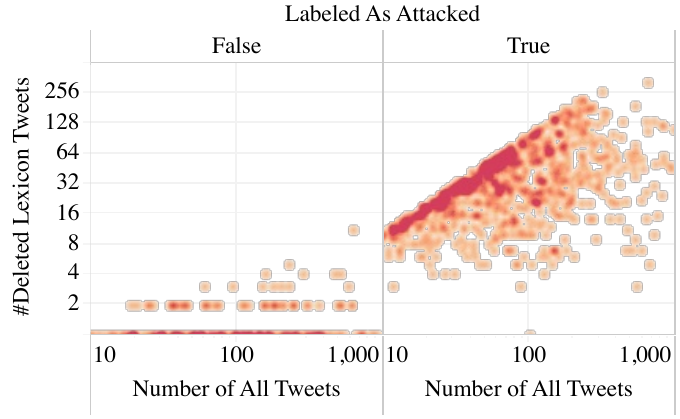}
    \caption{The number of deleted tweets classified as lexicon and number of all tweets per trend labeled as attacked (right) and other (left). Four deleted tweets classified as lexicon clearly separate the two classes.}
    \label{fig:separation}
\end{figure}

\Figref{fig:separation} shows the results of our lexicon classifier on the deleted tweets. It can separate the positive and negative cases in most cases. The classification task is then to account for the few false positives and negatives.

\descrplain{Classification} We sort the trends by date and use the first 80\% as training data. The test data starts from trends in February 2019. The training set contains 648 positives and 2,756 negatives while the test set contains 195 positives and 656 negatives. A simple decision tree that checks if there are at least 4 deleted lexicon tweets associated with a trend and if more than 45\% of all lexicon tweets are deleted achieves a 99.7\% 5-fold cross-validation score, 100\% precision, 98.9\% recall, and 99.4\% F-score. The classifier can achieve such good results because it is classifying a very specific pattern that came to be due to a vulnerability in Twitter's trending algorithm. It is no surprise that the attackers have not changed their attack method since their current method is already very successful. Note that 4 lexicon tweets in the 1\% sample maps to roughly 400 lexicon tweets in reality, a clear anomaly considering that lexicon tweets are rare. To support our analysis, we further analyzed features of the attacked trends which were not needed for classification but provide insight into the anomalousness of attacked trends such as entropy (see Appx.~\ref{sec:features}). We also provide our additional experimental results in which we come up with a lexicon-agnostic classifier but did not use it while classifying the past instances of the attack (see Appx.~\ref{sec:experimental}).

This classifier found 32,895 trends (19,485 unique keywords) associated with at least one attack between June 2015 and September 2019. Most were created from scratch (astroturfed) but very few were promoted after they reached trending (see~\Secref{sec:attack_analysis}). We refer to these as \emph{attacked trends} for the remainder of this paper.

\descrplain{Classification of Astrobots}
Classifying any user who posted a tweet containing an attacked trend with a lexicon tweet deleted within the same day as an astrobot yields 108,682 astrobots that were active between June 2015 and September 2019. 44.9\% of these users do not remain on the platform as of July 2020. Through the users/show endpoint of Twitter API, we found that \textbf{27,731} of these users are \textbf{suspended}. For the rest (21,106), we are given a user not found error (code 50). Those users may be deleted by the account owners. We leave a fine-grained classification of astrobots to future work. 

\descrplain{Other Countries} We manually examined temporal activity (i.e. the number of tweets and deletions per minute) associated with non-Turkish trends with more than 10 deletions but did not find any positive example. We additionally built a lexicon-agnostic classifier and ran it on all hashtags contained in non-Turkish hashtags but failed to find positives that we could reliably assess as \name astroturfing. See Appx.~\ref{sec:other_countries} for details on this analysis. Thus, the remainder of the paper will focus on ephemeral astroturfing attacks on Twitter trends in Turkey.

\section{Attack Analysis}
\label{sec:attack_analysis}
In this section, we analyze the trends associated with the attacks to first answer if the attacks cause or just promote the trends. We then measure the success of the attacks using various metrics. We also examine the other tactics the attackers may have employed by studying the time of the trends and the tweets' location information. Lastly, we show an anomaly in the volume field provided by Twitter which shows how many tweets talk about the associated trend and discuss what it may signify.

Part of this analysis requires a dataset of trends that contains their exact time and ranking. We were unable to find such a historical dataset; however, we collected a detailed dataset of the top 50 trends in real-time from Turkey and the world between June 18, 2019, and September 1, 2019, by sending API requests to Twitter every 5 minutes. We name this dataset the \emph{real-time trends} dataset. 

\subsection{Causation or Promotion?}
Our initial observation which we build our classification method upon is the enormous number of lexicon tweets being created and subsequently deleted before the new keywords reached the trend list. As our retrospective dataset does not contain the exact time a trend reaches trending, we could not use this information in our classification and only classify if a trend is attacked at some point. We now show that for the majority of the trends, the attacks are the only activity before the target keyword becomes trending, and thus, attacks cause the trend.

Using the real-time trends dataset, we first collect all tweets associated with each trend from the retrospective dataset, before the first time the trend reaches the top 50 trends list until the first time they dropped from the top 50 trends list. Twitter states that the trending algorithm shows what is popular now~\cite{trendsfaq} which means that they take recency of tweets containing a trend as input. We did not see any major difference in the results when we only consider recent tweets, i.e. tweets created within an hour, and thus show results without accounting for the recency. We later found that this was because attack tweets were generally very recent, created within five minutes before the target keyword becomes trending (See \S\ref{subsec:howfast}).

\begin{figure}[!htb]
\begin{center}
\includegraphics[width=\linewidth]{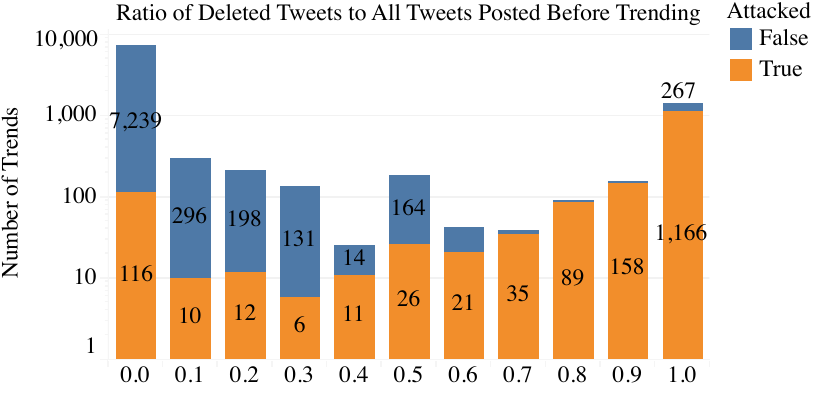}
\caption{The histogram depicting the ratio of all tweets that are created and deleted to all tweets created before the trend enters the list. This ratio is overwhelmingly high for attacked trends while it is zero for the majority of non-attacked trends.}
\label{fig:deleted_before}
\end{center}
\end{figure}

\Figref{fig:deleted_before} shows the ratio of tweets deleted to all tweets for each trend. Strikingly, the attackers delete all their tweets \textbf{even before the target keyword reaches trending} in n = 1166 / 1650 (70.6\%) cases. This demonstrates that Twitter's trending algorithm does not account for deletions. The attackers likely delete quickly because they aim to provide their clients with a clean slate once the keyword reaches trending. Additionally, the attackers may want to hide the fact that this is a fake trend from the public since people can see the attack tweets once the target keyword reaches trending by clicking on the keyword on the trends list. Very few non-attacked trends have a high percentage of deletions. These trends have less than four tweets found in the retrospective data and as such, they are either false negatives or noise. 

For 90.6\% of the attacked trends, the tweets deleted within the same day make up at least half of the discussions. Our further analysis yields that these deleted tweets are indeed lexicon tweets. We examined the data of 155 attacked trends in which deletions make up less than 50\% before they reach the trends list. We found that 24 attacked trends did not have any lexicon tweets, suggesting that they may be attacked at another time they were trending. For 56 trends, the lexicon tweets are deleted after the trend entered the list. Only for 37 trends, there were less than 4 lexicon tweets before the trend enters the list and there were many more lexicon tweets posted after the trend reached the list. These trends initially reached the list in a lower rank (median 32.5) but their highest ranks are in the top 10 with only 2 exceptions, suggesting that attacks promoted these trends rather than creating from scratch. The rest of the trends have prior lexicon tweets but also had some sort of other discussions. Thus, for at least 90.6\% of the cases, the attacks create the trends from scratch while for only 3.7\% of cases we can argue that the attacks are employed to promote a trend. 

\subsection{Success Metrics}

\subsubsection{Binary Success}
For measuring success, we begin with the simplest metric: does the target keyword reach trends? We detect unsuccessful attacks by running our classifier on tweets associated with keywords that were not trending on that day. If the classifier yields positive, that would mean there was an ongoing unsuccessful attack. We only use hashtags as a proxy for trend candidates as it's computationally expensive to run our classifier on every n-gram in the data. We collect all hashtags and their tweets between June 2015 and September 2019 from the retrospective dataset. We found only 1085 attacked hashtags that did not make it to the trends on the same day or the day after. 169 of those hashtags trended another day. As the number of trends that are hashtags since 2015 June is 21030, we estimate that attacks are successful by 94.8\% of the time. However, our results may be biased towards the attacks that employed sufficiently many bots with, which our classifiers can produce a positive estimate. %

We consider two main reasons that an attack fails: 1) the attack is not strong enough to be selected as a trend (at least not stronger than the signals associated with organic trends) by the trending algorithm and 2) the attack is filtered by Twitter's spam filters. In the former case, per our attack definition, the failed attack may have fewer posts than the other candidate trends ($|\tweetset| < \numtweetsthreshold$), or the time window of the correlated tweets may be too wide ($max(\createdset) - min(\createdset) > \timeframethreshold$). In the case where the attack is filtered by Twitter's spam filters (as in~\cite{cummings}), we observe that some attacks include phone numbers (e.g. \textit{\#Whatsapp0***x***x****} *'s are digits), profanity (i.e. \textit{\#M****G*t}) or words related to porn (e.g. \textit{\#Pornocu\"{O}******}, which target an individual claiming he likes porn). There are also cases where the attackers made obvious mistakes e.g. they intended to push ``A\u{g}a\c{c} Bayram{\i}" (Tree Fest), but ``A\u{g}a\c{c}" (Tree) trended, or they typed the target keyword twice and tried to push \textit{\#53Y{\i}ll{\i}kEsaretYard{\i}mc{\i}Hizmet-\#53Y{\i}ll{\i}kEsaretYard{\i}mc{\i}Hizmet} and failed because the keyword was too long or it has two hashes. Since the number of unsuccessful attacks is too low and we are limited to only 1\% of tweets, it is nontrivial to find exactly why each attack was unsuccessful.

\subsubsection{Rank} 
\label{subsec:howhigh}
\begin{figure}[h!]
\begin{center}
\includegraphics[width=\columnwidth]{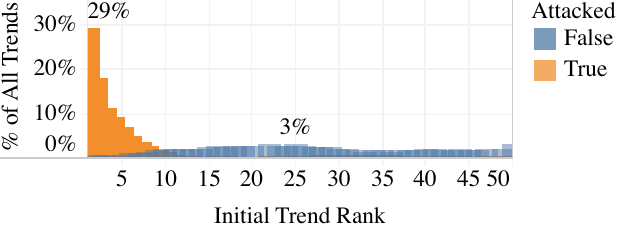} 
\caption{Histogram of the trends' initial rank for the attacked trends versus non-attacked trends. Attacked trends' usually rank in the top 5 with the majority ranking 1\textsuperscript{st}.}
\label{fig:ranks}
\end{center}
\end{figure}

Another measure of success and an indicator that the attacks cause or help trends tremendously is the attacked trends' ability to climb higher and faster than other trends. \Figref{fig:ranks} shows that the rank of trends when they reach the trends list for the first time follows a nearly uniform distribution. However, for the attacked trends, almost all rank in the top 10 with the majority ranking in the top 5 initially. This also shows that attackers' goal is to make the target keyword visible on the main page of Twitter or explore section on its app.

\definecolor{barpink}{HTML}{d48fc6}
\definecolor{bargreen}{HTML}{2b6f39}

\subsubsection{Speed} 
\label{subsec:howfast}
\begin{figure}[!htb]
\begin{center}
\includegraphics[width=\columnwidth]{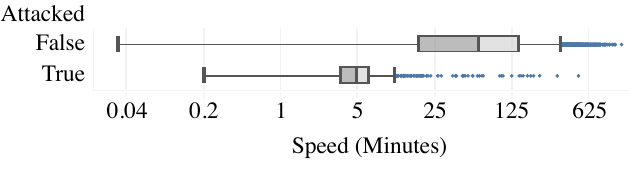}
\caption{The speed of keywords reaching trending. Most of the attacked trends reach trending around just 5 minutes, very fast when compared to other trends (median: 63 minutes).}
\label{fig:howfast}
\end{center}
\end{figure}

In addition to reaching higher, attacks also make a keyword reach trends faster than other trends. To measure this, we subtract the median time of tweets posted before the associated keyword reaches the trends list for the first time from the time it reaches trends which we name the \textbf{speed} of a trend. \Figref{fig:howfast} shows that the speed of attacked trends is much higher and concentrated around 5 minutes which amounts to the time Twitter refreshes the trends list. This suggests that the attackers do not even start some sort of discussion before the target keyword, but just attack with enough bots to make it reach the trends suddenly. 

\subsubsection{Duration} 
\begin{figure}[ht!]
    \centering
    \includegraphics[width = 0.9\linewidth]{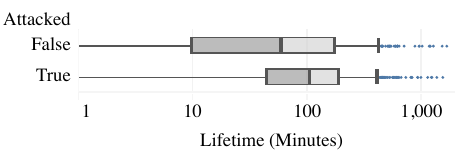}
    \caption{Lifetime of top-10 non-attacked trends (top) versus attacked trends (bottom). Attacked trends tend to stay longer (median: 105 minutes) in the trending list when they initially enter the trends list even when compared to other top 10 trends (median: 60 minutes).}
    \label{fig:trendlife2}
\end{figure}

Another measure of how well an attack succeeds is how long the attacked trends stay in the trends list. The attacked trends stay in the trends list for longer even when compared to non-attacked trends that also entered the trends in the top 10, as \Figref{fig:trendlife2} shows. The initial attack's strength may influence the length of the trend. However, additional actions may play a role in influencing the length of the trend. The attacks may be combined with an organic or an inorganic campaign or a mixture of two (as in \#İstanbulunUmuduİmamoğlu in \Figref{fig:ex}) or may capture the attention of the public which discusses the trend for an extended amount of time (as in \#SuriyelilerDefolsun in \Figref{fig:ex}) or the trend is promoted by subsequent attacks (as in \#SamsununAdresiMacellanCafe in \Figref{fig:ex}).

\subsubsection{Impact on Trends}
\begin{figure*}[ht!]
    \centering
    \includegraphics[width=0.9\textwidth]{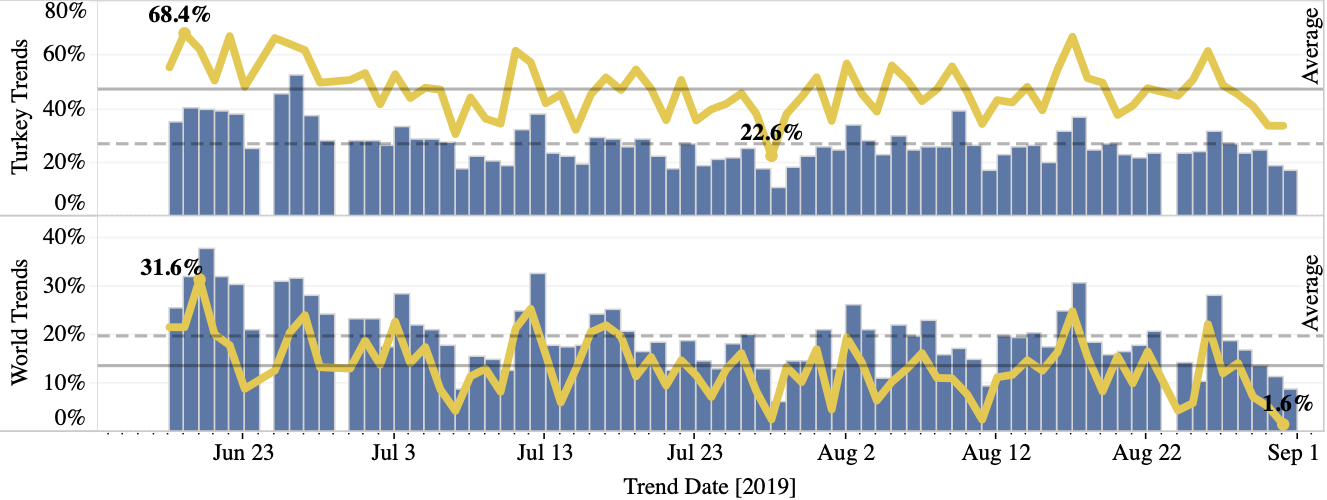}
    \caption{Percentage of the attacked trends reaching the top 10 (bars) and the top 5 (lines) trends in Turkey (top) and the world (bottom.) per day. 
    The daily average of attacked trends positioning themselves in the top 10 trends in Turkey is 26.7\% while this value goes high as 47.5\% for the top 5. The highest value is 68.4\% on 19 June 2020, four days before the Istanbul election rerun and the minimum value is 22.6\%. The daily average of attacked trends positioning themselves in the top 10 global trends is 19.7\% and 13.7\% in the top 5, maximum 37.9\%, and 31.6\% respectively.}
    \label{fig:attackperday}
\end{figure*}

Now that we have shown how successful the attacks are individually, we estimate the prevalence of this attack in terms of the percentage of daily trends that are manipulated. To measure the prevalence, we record how many unique target keywords we know to be artificially promoted by \name astroturfing attacks per day and reached the trends list, and compare it to the total number of unique trends on the same day. From June 2015 to September 2019, we found 32,895 attacked trends, making up 6.6\% of our top 50 trends data since June 2015. However, this is likely an underestimation. First, because not all trends' data are found in the 1\% real-time sample. More importantly, as we observe in \S\ref{subsec:howhigh}, attacks only aim for the top trends because only the top trends are visible and would make an impact. Therefore, using our real-time trend dataset, we compute the percentage of the attacked trends to all trends positioning themselves in the top 5 and the top 10 trends. Figure \ref{fig:attackperday} shows the percentage of top trends that are attacked for the trends in Turkey (upper) and the world trends (lower), positioning in the top 10 (bars) and the top 5 (lines). The daily average of attacked trends reaching the top 10 is 26.7\%. This number goes as high as 47.5\% for the top 5, reaching the highest on July 19, 2019, to 68.4\%, 4 days before the June 23, 2019, Istanbul election rerun. Crucially, many of these keywords reached world trends. The daily average of attacked trends reaching the top 5 is 13.7\% reaching the highest 31.6\% while this number is 19.7\% for the top 10 trends with a maximum value of 37.9\%. 

\subsection{Tactics}
\subsubsection{Time of Day}
\begin{figure}[ht!]
    \centering
    \includegraphics[width=\linewidth]{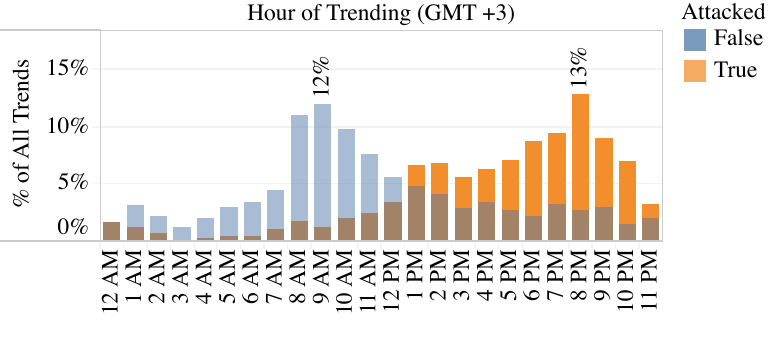}
    \caption{Percentage of keywords entering the trends list in a specific hour. The attacked trends enter the trends list mostly at night (Turkey time) while others enter in the morning.}
    \label{fig:when}
\end{figure}

We now turn to one of the tactics the attackers may be employing, sniffing the best time to execute attacks. For those trends which make to the top 10, the trends that are not associated with attacks generally enter the trend list in the morning while the attacked trends mostly enter the list at night as Figure \ref{fig:when} shows. The attackers may be choosing nighttime to maximize the binary success; they may be assessing the agenda of the day to decide on how many astrobots to employ, or whether to attack or not. It may be also because the organic trends tend to enter the trend list in the morning possibly due to news setting the agenda and creating competition. Alternatively, it may be due to maximizing the impact; the attackers may be considering the night hours as a better time to attack since people may be surfing on Twitter more at night and thus be more susceptible to attention hijacking.

\subsubsection{Location Field}
\begin{figure}[ht!]
    \centering
    \includegraphics[width=\linewidth]{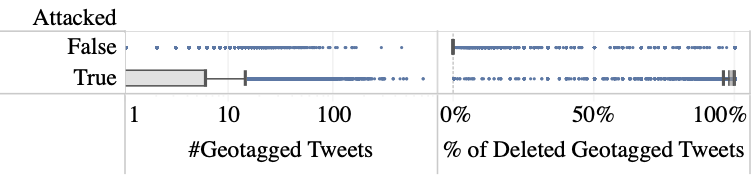}
    \caption{The number of geotagged tweets (left) and the percentage of geotagged tweets deleted to all geotagged tweets(right), per trend. Attacked trends have more geotagged tweets and the majority are deleted.}
    \label{fig:geotags}
\end{figure}

It is likely that attackers spoof locations to make the trend nationwide instead of in a single city. Additionally, the trending algorithm may be favoring trends with geotagged tweets or trends discussed in a wide range of locations. Similar behavior was reported in~\cite{safaksari} in which pro-government Twitter users organize a disinformation campaign against Istanbul's mayor about a water shortage in Istanbul but the tweets are posted from 16 different cities. To show this, we collect the geotagged tweets in the retrospective data, 285,319 tweets in total. Of the 285,319 geotagged tweets in the retrospective dataset, 77.63\% are associated with attacked trends even though their tweets make up 25.3\% of all tweets. 95\% of the geotagged tweets associated with attacked trends are deleted while this is only 14\% for other trends. \Figref{fig:geotags} shows the number of geotagged tweets and the percentage of deleted geotagged tweets to all geotagged tweets per trend. 

To verify these geotags are indeed fake, we tracked 5,000 users which we manually confirmed were astrobots, in real-time for one week. Out of the 3140 bots active at that time, 384 had at least two distinct geolocated tweets. We then compute the total distance between all of the points (in chronological order) in a 5-day span for each account. The average distance covered in one week by astrobot accounts was 24,582 km: a round trip from Istanbul to the capital, Ankara, 70 times.

\subsection{The Volume Of Trends} 
\begin{figure}[ht!]
    \centering
    \includegraphics[width=\linewidth]{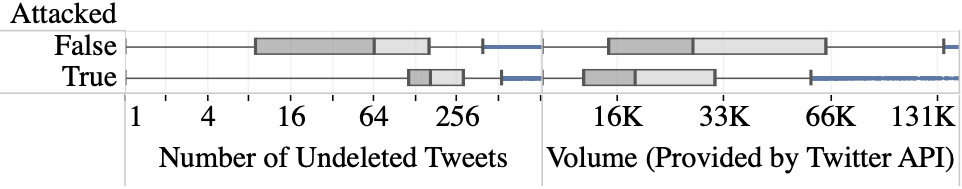}
    \caption{The number of undeleted tweets related to attacked trends and other trends vs the volume field provided by the Twitter API. While the former is higher for attacked trends (median is 166 vs 64 for other trends), the latter is higher for other trends (median is 27k versus 18k for attacked trend). This may mean that Twitter filters out the inorganic behavior associated with trends while computing the volume. The minimum volume is 10,000 likely because Twitter sets the volume to null when it is below 10k. }
    \label{fig:volundeleted}
\end{figure}

The Twitter API’s \textit{GET trends/place} endpoints both provide the trends and their \emph{volumes} which signifies the number of tweets discussing the trend as computed by Twitter’s internal tools. Though in reality the number of tweets posted to an attacked trend is higher than other trends, the \emph{volume} of attacked trends is lower compared to other trends, as \Figref{fig:volundeleted} shows. While the black-box nature of the volume field obscures the true reason, it may be that attacked trends were promoted by other bot-like accounts that Twitter discarded while computing the volume of tweets associated with trends.

\section{Account Analysis}
\label{sec:useranalysisshort}
In this section, we analyze the types of accounts the attackers employ. We sampled 1,031 astrobots that were employed in these attacks which were still active in March 2019. We inspected the profile and the recent tweets of the accounts and came up with categories based on the content and time of their tweets in an open-world setting. One author annotated all accounts and another annotated 100 to report inter-annotator agreement, which was K = 0.707 (substantial agreement.) The annotators established three categories that covered 947 accounts (92\%): 1) inactive (zombie) accounts, which are either dormant for years or have no persistent tweets but actively post lexicon tweets and then delete them (n = 304), 2) retweeter accounts, whose timelines are made up of only retweets (n = 157), and 3) accounts that appear to be human due to their \emph{sophisticated}, original, recent tweets (excluding retweets) and conversations with other users on the platform. We defined sophisticated as containing genuine sentences that convey meaning and have standard grammar (n = 486). 

We suspect that most if not all of the users from the latter group are compromised accounts, which were also reported by Turkish media~\cite{hurriyet,bbcturkce,cengizsemerci}. The most compelling evidence to support this is that the accounts' political orientations observed in undeleted tweets and deleted tweets are inconsistent. Pro-opposition hashtags such as \textit{\#Her\c{S}ey\c{C}okG\"{u}zelOlacak} (a candidate's slogan, \textit{\#EverythingWillBeGreat}) are the most prevalent and adopted by 104 users. However, the most prevalent hashtags among deleted tweets of this otherwise pro-opposition group of accounts are obvious spam advertising trend topic service and/or fake follower schemes. When we examine the hashtags in the deleted tweets, we find that they contradict the political views found in the other tweets: 43 tweeted the pro-government hashtag \textit{\#Erdo\u{g}an{\i}\c{C}okSeviyoruz} (\textit{\#WeLoveErdo\u{g}an}), and 19 tweeted with the anti-opposition hashtag \textit{\#CHP\.{I}PinPKKl{\i}adaylar{\i}} (\textit{\#PKKMembersAmongCHP\&\.{I}yiParty}) which claims that the opposition is aligned with terrorists. 

We also contacted and interviewed 13 users whose accounts appear to be compromised, using a non-Twitter channel when we were able to locate the off-platform (Twitter = 8, Instagram = 3, Facebook = 2). We informed the user that their account was being used in a botnet and verified that their account was compromised, with the attacker taking partial control. The users either did not know they were in the botnet, or they were helpless and did not think it was a big enough problem to address since the attackers only use them to astroturf trends and quickly delete their tweets. 


On June 12, 2020, Twitter announced that they suspended and published the data of 7,340 fake and compromised accounts that made up a centralized network~\cite{twitterannouncement}. The accompanying report by the Stanford Internet Observatory claimed that the accounts, among other tactics, employed astroturfing, aimed at pressuring the government to implement specific policy reforms.”~\cite{fsi}. The report did not mention fake trends created by the attack we describe here and nor their prevalence. To show that part of these accounts removed on that occasion were indeed astrobots, we cross-referenced these accounts with those in our retrospective dataset. We found an overlap of 77 accounts which we manually identified as astrobots as they tweeted lexicon tweets. Of these, 27 had lexicon tweets that were published by Twitter and publicly accessible. We examined the non-deleted tweets of all 77 accounts to identify their purposes. Only 5 of these accounts appeared to be pro-government while 25 exhibited bot-like behavior since they were only employed to promote hashtags on policy reforms. Eight users were openly anti-government while the rest appeared to be apolitical. This further backs up our claim that some of the astrobots are non-pro government accounts that are compromised, as this is also how Twitter framed the accounts they suspended in this dataset. There are likely many more compromised accounts astroturfing trends in this dataset, but we cannot identify more without access to any deleted tweets, which Twitter did not share.


We combined this data with the deleted tweets found in our retrospective data and identified 77 astrobots tweeting lexicon tweets. Of these bots, 27 were identified through the data Twitter provided since the attackers did not delete these users' lexicon tweets. We examined the non-deleted tweets of these 77 accounts to identify their purpose. Only 5 appear to be pro-government while 25 have bot-like behavior, as they were only employed to promote hashtags on policy reforms. Eight users were openly anti-government while the rest appear to be apolitical. Our findings are in line with Twitter's, which announced that they had suspended non-pro government users that were compromised. There are likely many more compromised accounts astroturfing trends in the dataset, but we were not able to identify more. Since Twitter did not share the deleted tweets, we needed to rely on the 1\% sample for deletions.

\section{Attack Ecosystem}
\label{sec:trendanalysis}

So far, we have claimed that the goal of the attack is to reach the trends list. However, if we take a step back there's a more important question to ask: ``\emph{Why} do the attackers want to reach the trends list?'' In this section, we analyze the trends themselves and the ecosystem that supposed the attack to uncover the motivations of attacks. 

\subsection{Topical Analysis of Trends}

To understand what topics the attackers promote, we first perform a qualitative analysis on the topics of the attacked trends. We collected all 6,566 unique astroturfed keywords that trended in 2019 and labeled them according to which specific group (e.g. political party) that each trend promoted if any. We also annotated a supercategory for different types of groups. Two annotators labeled 3,211 keywords, one using network information if available (i.e. if two keywords are promoted by the same set of users) and the other using only manual examination. The manual examination consisted of reading the keyword itself or searching for it on Twitter and/or Google for the context. The annotator agreement was K = 0.78 (substantial agreement). The remaining 3,355 were annotated by only one annotator due to the absence of network information. The resulting supercategories with the group annotations in their descriptions are the following: 

\descrplain{Illicit Advertisement (n = 2,131):} Trend Spam, i.e. trend topic (TT) services, or the fake follower schemes (n = 259), betting websites (n = 1421), a local social media platform (n = 27), a logo design service (n = 20), illegal football streaming websites (n = 24) and others (n = 380). Advertisements often promote the same entity multiple times using slightly changed keywords. This may be because the attackers believe that the trending algorithm favors new keywords. We also observed that these trends are not usually tweeted about by real users. The account of the advertised entity was the first, and in some cases, the only, account to include the trending keyword in a tweet, i.e., attackers push a gambling website to trends, then the betting website's Twitter account uses the new trend and becomes visible in the “Popular Tweets" and “Latest Tweets" panels on Twitter.

\descrplain{Politics (n = 802):} Political slogans or manipulations in support of or against a political party or candidate. Pro-AKP keywords in support of the Turkish government (n = 348) and those that target the opposition (primarily CHP) negatively (n = 124) are the majority. There are also slogans in support of the main opposition party and its candidates (n = 118), other parties (n = 42), or targeting AKP (n = 20). The rest are keywords related to politics but not political parties. 

\descrplain{Appeal (n = 1,219):} Appeals to government suggesting some policy reforms, as in~\cite{fsi}. These state the demand of the client either in a camelcase form that makes up whole sentences (e.g. MrErdoganPlsGiveUsJobs) or is a very long hashtag (e.g. \#JobsTo5000FoodEngineers). The demands are for jobs (e.g. food engineers, contracted personnel, teachers, etc.) (n = 730), for pardoning prisoners (n = 157), for military service by payment (n = 54), and on other profession-related issues. Some of the demands are heavily opposed by the government, which suggests that attackers do not always work in favor of the government.

\descrplain{Cult Slogan (n = 592):} Trends that are about various topics but are all sourced from either the Adnan Oktar Cult (n = 474), Furkan's Cult (n = 105), or the Tevhid cult (n = 13), as the users campaigning using the corresponding trends explicitly state they are associated with the cult. All of the cults' leaders were arrested and some of the trends demanded the release of their leaders. Other trends include promoting the cults' views, e.g. spreading disinformation about the theory of evolution and the LGBT community. Turkish media has reported that Adnan Oktar and Furkan cults manipulate trends using bots~\cite{medyaradar,habernediyor}.

\descrplain{Boycott (n = 92):} Appeals to people to boycott Metro and MediaMarkt (n = 45) or other companies (n = 47).

\descrplain{Miscellaneous (n = 1,730):} Trends that are about any topic including social campaigns, TV shows, names of individuals, or random slogans. As they do not have interest groups, they may have been purchased by people who do not have the intention to campaign and may not be involved in multiple attacks. This corroborates that attacks are a business model with a wide range of clients, which is also reported by the Turkish news media~\cite{hurriyet,cengizsemerci,habericin}.

\subsection{Astrobot Network}
\label{sec:ecosystem}

\begin{figure}[ht]
    \centering
    \includegraphics[width=\linewidth]{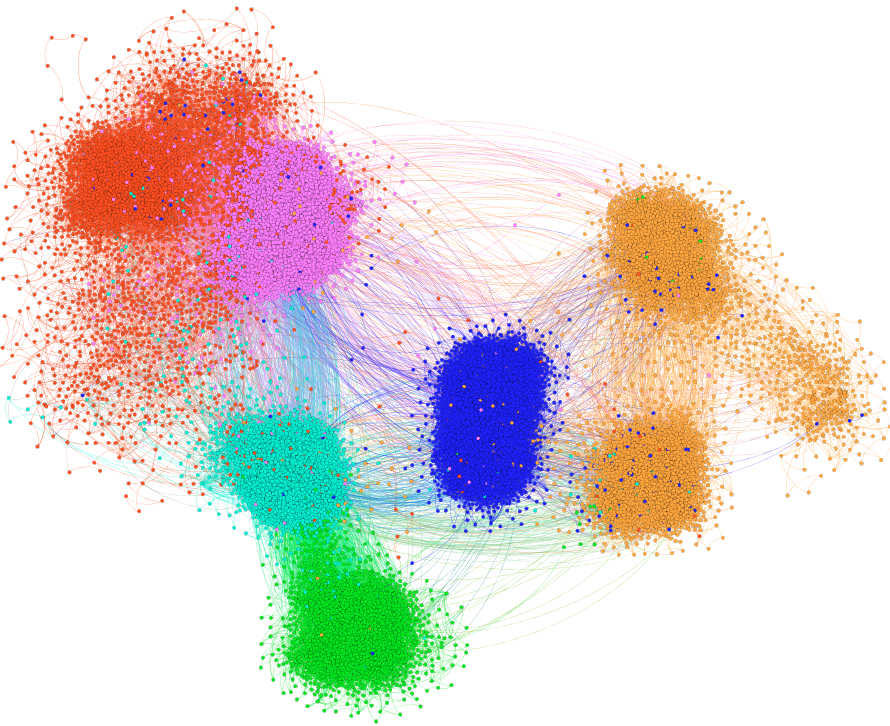}
    \caption{The astrobot network visualized in OpenOrd~\cite{martin2011openord} layout using Gephi~\cite{bastian2009gephi}. Colors indicate the communities obtained by the Louvain method~\cite{blondel2008fast}. The attackers lost control of the green and cyan communities by February 2019 while the remaining communities still participate in the attacks by September 2019. Spam trends that promote the fake follower service to compromise more users or promote the top trend service are mainly sourced from the blue community which has a central position in the network.}
    \label{fig:astrobot_network}
\end{figure}

As our attack model indicates, each trend is promoted by a set of \textit{astrobots}. The same set of bots can promote any trend as long as the bots are still controlled by the attackers. Thus, a set of bots consistently attacking the same trends are assumed to be controlled by the same attacker. Then, the same set of bots promoting keywords related to conflicting groups (e.g. opposing political parties) would indicate that the attacks are not executed by the interested parties, but that the attacks are provided as a service to different clients. To explore this, we extract and characterize communities of astrobots by analyzing their topical and temporal activities. The latter provides insights into how Twitter may be defending the platform.

We build the astrobot graph network in which the nodes are accounts and the edges indicate that the accounts participated in the same attack (both posted a deleted lexicon tweet containing the trend). This network had 33,593 users active in 2019, 71.6\% of which were still active as of July 2020. Surprisingly, the intersection of the set of users promoting the trend and not deleting the tweets (147,000 users) and the set of astrobot accounts was only 817, suggesting that the astrobots' only function is to push keywords to trends and stay idle otherwise. This is likely part of the stealthy nature of \name astroturfing attacks; the attackers do not want any of their bots associated with the campaigns they promote, so they do not employ them for non-\name astroturfing attack tweets. Instead, the clients outsource pushing trends to attackers and then execute any other activity themselves.

From the users active in 2019, we removed those who attacked only once to remove noise, leaving 21,187 users. We performed community detection using the Louvain method~\cite{blondel2008fast}. The Louvain method greedily optimizes modularity, which ensures nodes that have strong connections are in the same community. Thus, astrobots consistently attacking the same trends will be in the same community. We found 6 communities, with modularity 0.711. We name these communities by the coloring scheme: green, cyan, blue, orange, pink, and red. \Figref{fig:astrobot_network} shows the resulting network. Table III shows the number of trends and users and the percentage of users that remain on the platform by July 2020 within each community, as well as any pattern(s) we found. We now describe the temporal and semantic patterns the communities follow in detail.

\begin{table}[ht]
\centering
\label{tab:astrobot_statistics_main}
\caption{Statistics of the communities. Persist denotes the percentage of users not suspended or deleted within the community as of July 2020. Summary refers to the pattern(s) that characterize(s) the communities.}
\resizebox{\linewidth}{!}{%
\begin{tabular}{lccccl}
\textbf{Community} & \textbf{Users} & \textbf{Persist} & \textbf{Trends} & \textbf{Activity} & \textbf{Topic}                                              \\ \hline
Green     & 2,079  & 81\%         & 291    & 1/19 - 2/19 & Misc. \\ 
Cyan      & 3,701  & 78\%         & 839    & 6/18 - 2/19 & Appeal (Pardon)   \\ 
Blue      & 4,845  & 70\%         & 1,043   & 1/19 - 9/19  & Ads (Spam)            \\ 
Pink      & 4,719  & 74\%         & 2,422 & 3/19 - 9/19  & Ads (Betting)            \\ 
Red       & 2,627  & 73\%         & 941  & 3/19 - 9/19  & Various             \\ 
Orange    & 3,216  & 71\%         & 913  & 4/19 - 9/19  & Cult (Furkan)                        \\ 
\end{tabular}%
}
\end{table}

\begin{figure}[ht]
    \centering
    \includegraphics[width=0.9\linewidth]{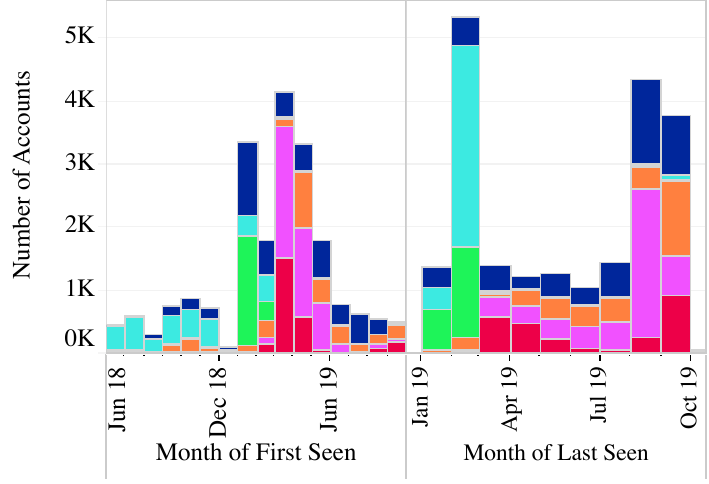}
    \caption{The time the accounts are first and last seen attacking. The users from the cyan community were active even before 2019 while the rest of the community became active in 2019. Accounts in the green and cyan communities appear to discontinue attacking in early 2019.}
    \label{fig:firstlastseen}
\end{figure}

\descrplain{Temporal Activity} Studying temporal activity is key to learning how many networks of astrobots are active at a given point in time and how quickly Twitter addresses the coordinated activity. \Figref{fig:firstlastseen} shows the first and last times the astrobots were found to be participating in an attack which gives us a rough idea of the active times of their respective communities.

\begin{figure}[ht]
    \centering
    \includegraphics[width=0.8\linewidth]{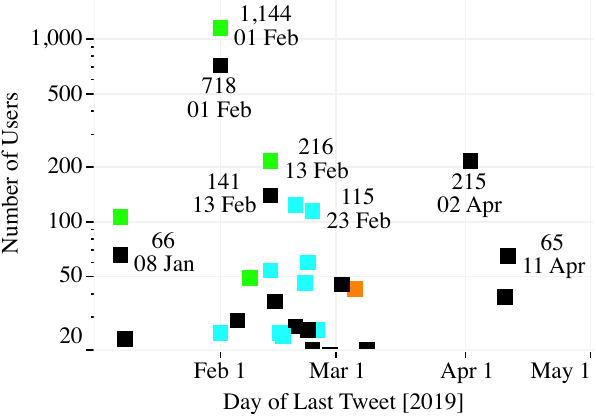}
    \caption{The date of last not deleted tweet of each account per community. Accounts shown in black are not assigned to a community. The huge spike in accounts that last tweeted on February 1 and never since may show that attackers lost control of these accounts, although the accounts are not suspended.}
    \label{fig:unsuspended_comm}
\end{figure}

We found that the cyan community is the only community in which the majority of the accounts (79\%) were actively participating in the attacks before 2019 while the rest became active in mid-2019. Exceptionally, the green community's users were active since January 2019. The green and cyan communities stopped attacking in February 2019, however, most of the accounts in these two communities remain on the platform despite not participating in recent attacks. All users that remain on the platform in the green community and half of such users in the cyan community last posted an undeleted tweet in February, as \Figref{fig:unsuspended_comm} shows. Precisely, 1,887 users became dormant on February 1, 2019. 23\% of the users in the green community and 6.7\% in the cyan community last tweeted a lexicon tweet in February, none of which were deleted, suggesting that the attackers could not or did not delete the lexicon tweets from their final attack using these communities. The other half of the users in the cyan community remained on Twitter as of July 2020 but did not participate in an attack after February. This suggests that the attackers may have lost control of the accounts either due to internal problems or because Twitter imposed a session reset or a location or IP ban on the accounts.

The fact that two of the communities were inactive by early 2019 and three new communities then became active indicates that the attackers replaced the cyan and green communities with new ones. Interestingly, while the majority of the creation dates of the accounts in the other communities are from 2016, 62\% of accounts in the green community were created between 2012 and 2014 even though they did not become active in any attacks until January 2019. Attackers may have bulk purchased and/or bulk compromised these accounts and were detected by Twitter quickly and taken down. The rest of the four communities were still participating in attacks as of September 2019. This indicates that there are four databases of astrobots owned by at most four different attackers. 

\begin{figure}[ht]
    \centering
    \includegraphics[width=\linewidth]{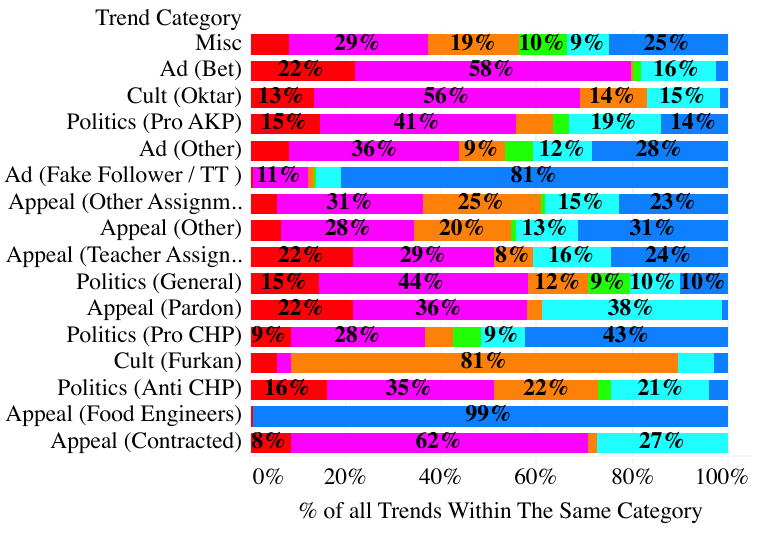}
    \caption{The trends according to topics (those with at least 100 trends) and the astrobot community the trends are promoted by. Some interest groups such as contract employees are merged into one.}
    \label{fig:semanticdist}
\end{figure}

\descrplain{Topical Activity} We now analyze the interplay between attackers and clients by analyzing the topical categories of the trends and the astrobot communities promoting them. \Figref{fig:semanticdist} shows the distribution using the topics from the previous subsection. Except for the green community, in which 60\% of trends were labeled as miscellaneous, no community was dominated by a topic and/or group. Some topics were mostly or uniquely associated with one community, suggesting that groups promoting those topics only collude with one attacker, although the same community promotes other topics as well. 

The majority of the bet related ads (80\%) and Oktar's cult slogans (68\%) were in the pink and red communities. Most (80\%) spam ads promoting fake follower schemes and trend topic services were in the blue community. The fact that this community is central to the whole network suggests that the attackers controlling this group provided users and trends for other groups. Food engineers appealing for jobs were also almost uniquely associated with the blue community, while cult slogans related to Furkan were associated uniquely with the orange community. Political trends were dispersed throughout the communities and political parties often shared the same community with their rivals. For instance, the blue community contains the highest number (80) of pro-CHP (the main opposition) trends but also has 80 trends in support of its fierce rival, AKP. Similarly, 40\% of the pro-AKP trends are associated with the pink community but the pink community has also 15 pro-CHP trends. This further corroborates that the attackers are independent of the parties and provide fake trends as a service. 

\section{Security Implications and Mitigation}

\label{sec:defenses}

\Name astroturfing attacks principally have an impact on users and the platforms that they attack in terms of (i) platform integrity, (ii) account security, and (iii) attack attribution. It also has a tremendous impact on data integrity in data science studies. We discuss further implications to security research and propose countermeasures.

\descrplain{Platform Integrity}
Systematic attacks on popularity mechanisms compromise the integrity of the mechanism and the platform. On Twitter, users expect to see legitimate popular content, so when they are shown content that is not popular, they no longer trust that what is shown on the Twitter trends list is actually popular. As with many systems, when the authenticity of a component is compromised, trust in the entire system diminishes, e.g. the price of bitcoin falls after prominent exit schemes. If Twitter trends fails to reliably display authentic trends, trust in trends and Twitter as a whole is diminished. Twitter recently took steps to preserve platform integrity such as suspending accounts involved in coordinated inauthentic behavior, however, they have not addressed \name astroturfing attacks which are contributing to a loss of trust among the users affected by the attacks. 

\descrplain{Account Security}
\Name astroturfing reinforces the practice of selling accounts. Because astroturfing attacks attempt to mimic widespread popularity, they require a critical mass of accounts, they necessitate a black market for compromised and fake accounts. \Name astroturfing is unique in that it allows for the use of active, compromised accounts and not only fake accounts. As long as \name astroturfing remains effective, more compromised accounts will be needed to boost the target keywords. While it is challenging to disrupt these markets directly, e.g., via takedowns, they can be disrupted by removing the market demand, rendering fake and compromised accounts useless.

\descrplain{Attack Attribution}
Malicious online activities are often difficult to attribute to an actor, and astroturfing attacks are no exception. Organic campaigns that are launched by users can generally be attributed to a certain group, ideology, or event. However, in the case of ephemeral astroturfing, the actions of the adversaries are quickly hidden. This makes it possible for adversaries to conduct illicit activities including the promotion of scams and illicit businesses. Ephemerality makes it more difficult to attribute an attack to a specific group, while at the same time legitimizing the activity by making it seem as though the activity is the result of grassroots organizing.

\descrplain{Data Integrity}
Beyond astroturfing, data science studies often rely on the assumption that data is a static entity. This is especially the case in social media studies, where data is often collected ex post facto. Such an assumption should be taken very carefully, as social media posts and accounts can be deleted or suspended. If accounts or posts are deleted, then the dataset used for evaluation and analysis may be incomplete. In the case of \name astroturfing, we find that a \emph{critical} segment of the data may be deleted: the tweets that illegitimately created the popularity of a topic. Future analysis of a trend that does not consider deleted data may misinterpret how a topic became popular. For example, in September 2018 the trend \textit{\#SuriyelilerDefolsun} (\#SyriansGetOut) was pushed to the trends list using ephemeral astroturfing attacks, as shown in \Figref{fig:ex}. The hashtag attracted the primarily negative attention of the public after reaching trending. However, academic studies that use the hashtag as a case study~\cite{cengiz2020political, ozduzen2020post} or refer to it~\cite{kirkicc2018educational, ccakir2020turkiye} all attributed the hashtag to the public, completely unaware of the fact that the hashtag was trending due to bot activity, even going as far as to say that social media users launched the hashtag due to a fear of a new wave of mass migration~\cite{pekkendir2018discursive}.

\descrplain{Impacts on Society}
\Name attacks expose users to illegal advertisements, hate speech targeting vulnerable populations, and political propaganda. For example, \c{C}iftlikBank was a Ponzi scheme that claimed to be a farming company aimed at growing quickly and aggressively maximizing profits. They employed \name astroturfing attacks to promote themselves 29 times using slogans such as \textit{\c{C}iftlikBank TesisA\c{c}{\i}l{\i}\c{s}{\i}} (\c{C}iftlikBank Facility Opening) and \textit{\c{C}iftlikBank BirYa\c{s}{\i}nda} (\c{C}iftlikBank is one year old) which give the impression that it is a growing business. They did not trend organically until December 2017, and only then because they started to raise suspicions~\cite{hurriyetciftlikbank}. They attempted to counter this suspicion by using \name astroturfing to push \textit{\#\c{C}iftlikBankaG\"{u}veniyoruz} (\#WeTrustIn\c{C}iftlikBank) into trends. \c{C}iftlikBank's boss scammed \$129 million before escaping in March 2018~\cite{ahvalciftlikbank,bbcciftlikbank}.

Taxi drivers in Istanbul used \name astroturfing to protest Uber~\cite{reuterstaxiciler}. Some of the slogans aligning with their campaign were used to sow hate against the drivers, e.g. \textit{\#KorsanUberKapanacak} (\#PirateUberMustShutdown), \textit{\#R\"{u}\c{s}vet\c{c}iUber} (\#UberTakesBribes), and \textit{UberAra\c{c}SahipleriAra\c{s}t{\i}r{\i}ls{\i}n} (\#UberDriversShouldBeInvestigated). Other hateful messages targeted specific individuals demanding their arrest, e.g. \textit{\#Fet\"{o}c\"{u}KuytulTutuklans{\i}n }(\#ArrestKuytulHeIsATerrorist). Alparslan Kuytul is the leader of Furkan Cult and has an anti-government stance. Others spread hate speech and disinformation targeting vulnerable populations; the LGBT community was targeted at least 24 times by these attacks in 2019 with trends such as \textit{LgbtiPedofilidir} (LGBTisPedophilia) and \textit{DinDevletD\"{u}\c{s}man{\i} Sap{\i}kLgbti}, (PervyLGBT is enemy of religion and state). Occasionally, counter campaigns were launched by the attack targets, also employing \name astroturfing attacks, e.g., \textit{\#UberiSeviyoruz} (\#WeLoveUber) and \textit{\#HalkUberiSeviyor} (\#PeopleWantUber) were used to counter the taxi slogans. Additionally, people seemed to react to the prevalence of trends that appear to be sourced from Adnan Oktar Cult by astroturfing trends like \textit{\#Adnanc{\i}larMilletSizdenB{\i}kt{\i}} (Adnan Oktar Cult, people are sick of you,) and \textit{\#SizinA*kAdnanc{\i}lar} (Expletives directed at the Adnan Oktar Cult using abbreviated profanity). 

Politically motivated groups employed attacks for smear campaigns spreading disinformation and hate speech. During the 2019 local elections in Turkey, many pro-government groups astroturfed trends to target the opposition (e.g. \textit{\#CHPPKKn{\i}n\.{I}zinde}, which indicates that the opposition's party follow a terrorist organization) and particularly opposition candidate Ekrem \.{I}mamo\u{g}lu who eventually won the election and became mayor of Istanbul. Trends targeting the candidate involved slander asserting that he lied (e.g. \textit{\#EkrandakiYalanc{\i}}, which means “Liar on TV”) and that he stole votes. The most popular astroturfed trend on this issue, \textit{\#\c{C}\"{u}nk\"{u}\c{C}ald{\i}lar} ("Because They Stole"), was explicitly organized and pushed to the trends by pro-government groups~\cite{turkishminuteekrem} and joined by the rival candidate Binali Y{\i}ld{\i}r{\i}m~\cite{aaekrem}. Ekrem \.{I}mamo\u{g}lu condemned the campaign~\cite{bianetekrem}. After, the Supreme Electoral Council decided to rerun the elections but did not state there was any ballot ringing involved; Binali Y{\i}ld{\i}r{\i}m later stated he expressed himself in a colloquial language and the campaign was "an operation to create a perception".~\cite{bianetyildirim}.

Although many users are exposed to these trends, the extent of the impact is unclear as we cannot measure engagements with trends directly. Meanwhile, public engagement metrics such as the count of retweets and likes per tweet are open to manipulation. However, based on their appearance on other platforms, some astroturfed trends succeed in receiving the public's attention. For example, the mainstream media in Turkey framed many political slogans that trended due to \name astroturfing as grassroots organizing, e.g., \textit{\#ÇünküÇaldılar} (\#BecauseTheyStole)~\cite{ahaberekrem}, \textit{\#HırsızEkrem} (\#EkremIsAThief)~\cite{bbchirsizekrem}, and \textit{\#SüresizNafakaZulümdir} (\#IndefiniteAlimonyisTorture)~\cite{yeniakitnafaka}. Users also posted these slogans on Ekşi Sözlük, a local social media platform where users post entries for terms, because ``the public discusses them." \textit{\#ÇünküÇaldılar} received 352 entries and \textit{\#AtatürkAtamOlamaz} received 145 entries. Perhaps one of the most impactful \name astroturfing attacks was \textit{\#SuriyelilerDefolsun} (\#SyriansGetOut), which was astroturfed on September 3, 2018, sparking widespread controversy. It was discussed extensively by the media~\cite{birgunsuriyeliler,medyaskopsuriyeliler}, academic works~\cite{cengiz2020political,ozduzen2020post,kirkicc2018educational,ccakir2020turkiye,pekkendir2018discursive} and other social media websites such as Reddit~\cite{redditsuriyeliler}, Ekşi Sözlük~\cite{eksisuriyeliler} (265 entries) and kizlarsoruyor.com~\cite{kizlarsoruyorsuriyeliler}.

\descrplain{For Security Research} Although we focus on one case of \name astroturfing, the methodology that we present in this study can be extended to other attacks on popularity mechanisms. All popularity mechanisms work through parsing content to determine what is popular at the moment, though for different mediums and with different definitions of popularity. Considering deleted activity or content as valid leaves open an attack vector, making \name attacks possible. Our results shed light on the problem of astroturfing, framed as an attack on popularity metrics, and how prevalent a problem it can become when left unchecked and unaddressed. 

\descrplain{Generalizability}
\Name astroturfing attacks can generalize to any platform where an algorithm determines trending or popular content and does not take deletions into account. However, this attack has yet to be explored on platforms other than Twitter. Traditional forums rank threads based on the time of the last reply, thus, spammers post comments to old threads to make them more visible, a practice called bumping~\cite{chen2015opinion}. However, forums generally account for deletions and rank the bumped threads lower when the bumper deletes their reply. Reddit considers the current number of upvotes and downvotes at the time it computes the ranking of threads and is therefore likely resistant to ephemerality, i.e. coordinated upvotes proceeded by removing those upvotes~\cite{redditranking}. Other possible vulnerable platforms include sites with reviews, like Amazon or the Google Play store, but so far no relevant public analysis of these platforms exists. This attack can also generalize to Twitter trends in any region. 

\vfill\null

\descrplain{Countermeasures}
Due to the use of active, compromised accounts, defenses against \name astroturfing attacks are inherently challenging. These accounts, whose owners are victims of the scheme, cannot simply be banned. If the attacks were being executed via a malicious application, Twitter could suspend access to the app, as in~\cite{businessinsider}, but in this case, tweets are posted from official Twitter apps (e.g. \emph{Twitter for Android}). Otherwise, \name astroturfing attacks fit an easily detectable pattern. We outline two main paths for defenses: detecting and inoculating. First, Twitter can implement a detection mechanism to prevent malicious tweets from being considered for Twitter trends, or even made visible at all. They can extend the detection method laid out in \Secref{sec:turkish} to find the tweets and accounts involved. Once a trend is found to be manipulated, it can be removed from trends or even prevented from ever reaching them. The second option is to render the attack useless. The fact that these attacks are successful implies that the Twitter trending algorithm does not consider the deletion of tweets. A simple defense, then, is to account for deleted tweets when computing trending topics. For example, the trending algorithm can track the tweets that contain the keyword and heavily penalize the trend's rank for each deleted tweet.

In addition to direct countermeasures, platforms can also work to ensure that even if a popularity mechanism is manipulated via deletions, that users can be aware of potentially suspicious behavior. On Reddit, for example, when a comment is deleted there is public evidence left behind that indicates that a comment was deleted. On Twitter, this translates to an indicator that a tweet that contained the trending keyword was deleted. In this way, when users click on a trend, they are not only shown tweets, but also a series of deleted tweets, which indicate that something suspicious has occurred. 

\descrplain{Limitations}
Study of \name astroturfing is limited to the platforms in which the content promoted by the popularity mechanism and the deletions are made available. While working with Twitter data, we are limited to only 1\% of tweets, as provided by Internet Archive's Twitter Stream Grab. Larger samples are not publicly available. This sample may not include attack tweets for every trend, so we may not be able to detect all attacked trends. Thus, we can only report lower bounds. We are also limited to local and global trends and are not able to analyze tailored (personalized) trends. The trending algorithm is black-box, and it is not reasonable to reverse engineer it using only a 1\% sample. Thus, we study the attack based on the behavior we observe in the data and only develop a classifier to detect one specific, ongoing attack instance.

\descrplain{Ethics and Reproducibility} This research was conducted using the Internet Archive's Twitter Stream Grab and trends data, so all data is public, and the study is reproducible. In addition, the IDs of the tweets and users annotated in this study as well as the annotated attacks are made available\footnote{https://github.com/tugrulz/EphemeralAstroturfing}. We acknowledge the ethical concerns with the honeypot account. To mitigate this, we signed up a newly created account which are normally filtered by Twitter's spam filters and minimized the amount of time that the account was active.

\bibliographystyle{IEEEtranS}

\bibliography{bib}

\begin{appendix}
\subsection{History of Astroturfing Attacks on Twitter Trends in Turkey}
\label{sec:historicity}
In this section, we introduce the astroturfing attacks prior to \name astroturfing attacks by a qualitative analysis similar to what we did in \Secref{sec:turkish}. These attacks are now obsolete but might still give insights on possible attacks in other contexts such as future attacks to Twitter trends in other countries or other social media platforms. 

We examine the first ten tweets associated with the trends to identify all kinds of astroturfing attacks and annotate the trends accordingly. We consider the trends that are initiated by tweets that are related to the trends they contain and appear to be written by a human as sophisticated. We annotate the trends with sophisticated tweets for which we can identify their origin as \textit{organic} and the others as \textit{related}. We consider the trends that are initiated by tweets which are not sophisticated but appear to be generated by a computer and follow a pattern that we can describe verbally as astroturfed. In total, we came up with 7 categories. Two annotators each labeled half of the 2,011 trends and one of them labeled an additional 200 randomly selected trends to reported annotator agreement, K = 0.607 for all 7 categories. 

\begin{enumerate}
\item{\textbf{Organic:}} The first tweets are sophisticated and they refer to a popular event or contain a hashtag that was started by a popular user(s), i.e. celebrities, politicians, or TV programs. (n = 765)

\item{\textbf{Related:}} The first tweets are sophisticated, but the trends are neither about an event nor contain a hashtag that was started by a popular account so the trends' origins are unclear. (n = 609)

\item{\textbf{Astroturfing --- Generic:}} The first tweets are not sophisticated and might be bot activity because the contents of the tweets are either i) only the target keyword (n = 93); ii) only the target keyword plus a random string or a number (n = 22); iii) low quality and repetitive comments shared by many trends, e.g. ``\#DonaldTrump let's make this trend" followed by "\#DonaldTrump oh yeah super trend" (n = 46). We call this type generic because we believe that it's easy to detect such attacks as the content injected is repetitive and can be applied to all trends regardless of the meaning they convey and/or the most of users self-declare that they aim to push the target keyword to trend list.

\item{\textbf{Astroturfing --- Premade:}} The first tweets are not sophisticated and likely bot activity because they are premade statements that are irrelevant to the trends. The trends are injected to arbitrary places in the tweets. e.g. famous quotes or full sentences irrelevant to the trends' meaning (i.e. Easy come easy \#DonaldTrump go. ) (n = 119) In some cases, a full paragraph from a story or a novel appear to be splitted into tweets, the target keywords are injected and the tweets are posted without even randomizing the order, making it easy to detect the pattern. The same behaviour is observed for the Russian bots in~\cite{golbeck2019benford} and the Saudi bots in ~\cite{Bbctrending} This tactic became dominant in 2014 but later fell out of favor in the same year. At the same time, news state that Twitter start to fight with the bots~\cite{robotlobbies} and impose restrictions on astroturfing attacks although they could not prevent them~\cite{hurriyet} which might suggest that the attackers changed their methodology and switch to ephemeral astroturfing attacks around that time.

\item{\textbf{Astroturfing --- Lexicon:}} The first tweets are clearly not sophisticated and likely bot activity because they appear to be sourced from a lexicon of words and phrases, i.e. "apple (fruit) to cycle elephant \#DonaldTrump trigonometry" (n = 182) 

\item{\textbf{Astroturfing --- Lexicon - Premade:}} Both premade statements and lexicon tweets are used for the same trend in different tweets. (n = 6) This might be due to the attackers testing the lexicon method. 

\item{\textbf{Other:}} The first tweets are either retweets or spammers hijacking trends by injecting an irrelevant message to top five trends, which we observe to be common in Turkish Twitter. These trends' origin is likely not captured by 1\% sample. (n = 166)

\end{enumerate}

\begin{figure}[ht!]
    \centering
    \includegraphics[width=\linewidth]{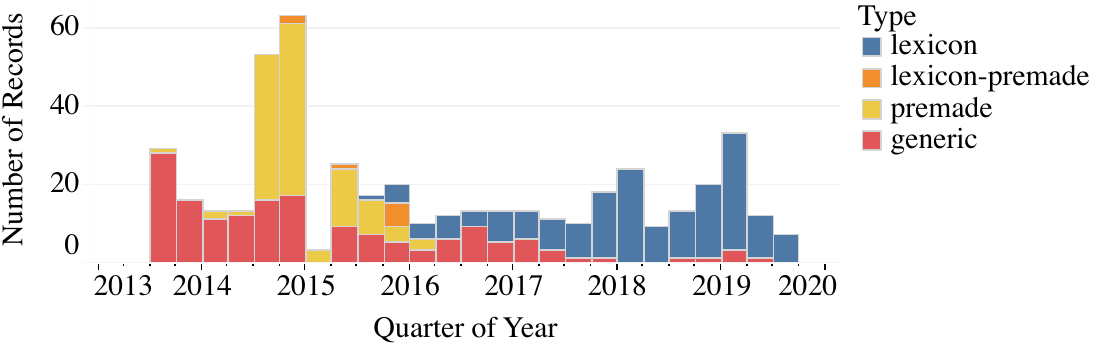}
    \caption{Frequency of different types of the attacks overtime (stacked). Premade and generic astroturfing attacks have fallen out of favor since early 2017 and lexicon astroturfing attacks have become the primary method for astroturfing. The gap in the first quarter of 2015 is due to missing data in the Internet Archive. The spike in 2014 coincide with presidential and local elections in Turkey.}
    \label{fig:procedure}
\end{figure}

\Figref{fig:procedure} shows the labeled categories by date for each of the manually labeled trends (recall that we randomly chose 1 trend per day to label). From our sample, we see that the premade and generic astroturfing attacks have fallen out of favor and lexicon astroturfing attacks now dominate the trends.

\begin{figure}[!htb]
\begin{center}
\includegraphics[width=\linewidth]{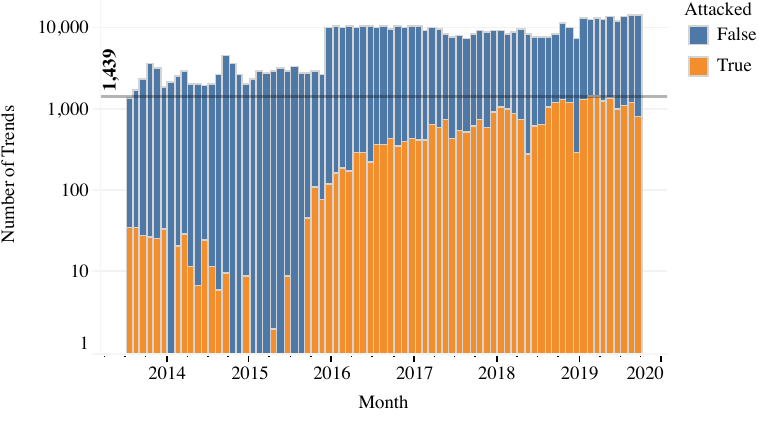}
    \caption{The start and spread of \name the astroturfing attacks on Twitter trends in Turkey. Attacked trends start to become widespread in 2016 and they peaked at the time of 2019 Turkish Local Elections (March) and 2019 Istanbul Election Rerun (June). The sudden drops in the attacks are due to missing data in archive.org's dataset while the increase in number of trends is due to the change in number of trends Twitter's API provide.}
    \label{fig:evolution2}
\end{center}
\end{figure}

Classifying all the \name astroturfing attacks using our classifier introduced in \Secref{sec:turkish}, we found attacks with lexicon tweets started in June 2015 and slowly spread throughout 2016 and was the dominant form of the astroturfing attacks by 2017 as \Figref{fig:evolution2} shows. We also found 251 trends predating June 2015 that have many deletions and tweets where the rules for lexicon tweets apply, but are of the generic type with random strings or low-quality comments e.g. "oh yeah this tag is nice". Examining these trends, we believe they are obvious spam as 106 of them have the keyword "takip" (follow) i.e. \textit{100de100GeriTakip} (Followback for 100\%) and appear to organize follow for follows while 41 of them have "tag" (short for hashtag) and they are variations of a slogan meaning ``Our Tag is the Best". 30 of them have the word "kazan" (win), i.e. \textit{\#CumaBuTagKazand{\i}r{\i}r} (Friday This Tag Will Make You Win). These trends do not qualify ephemeral astroturfing attacks since they do not follow the attack model we describe (either due to low rate of deletions or the creations and deletions' time window are bigger than attacks with lexicon tweets)

We observe that the attack tweets from different astroturfing attacks have the following attribute in common: they only engage with the target keyword. Therefore they do not include other entities within the tweets that Twitter acknowledges as metadata attributes such as profile handles (i.e. \@realDonaldTrump), hashtags (other than the target keyword) and urls. This phenomena is also observed in ~\cite{golbeck2019benford, Bbctrending}. Attackers might be using such tweets to pass spam filters of Twitter. We call tweets that only engage with the target keyword \textbf{single engagement tweets}.

\subsection{Analysis of Trends' Features at Scale}
\label{sec:features}
In this section, we further show that we captured the trends with attacks we described and not trends with an alternative behavior by analyzing the features which approximate the behavior of the attacks. Although they are not used by our classifier, we find them insightful in understanding the behavior of the attacks. We use all the data in the retrospective dataset. 

\begin{figure}[!htb]
\begin{center}
\includegraphics[width=\linewidth]{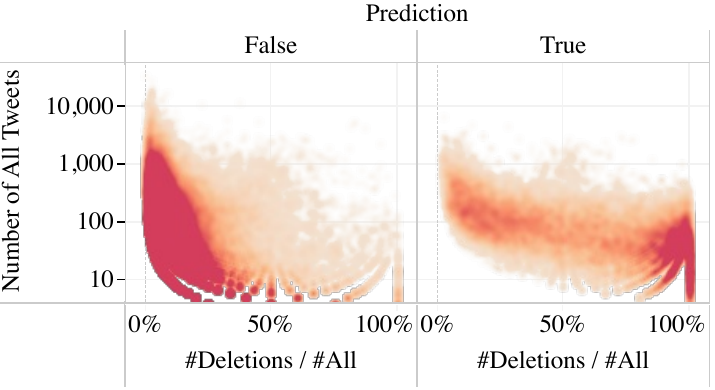}
    \caption{Percentage of deletions of trends associated with attacks and other trends. Attacked trends' deletion rate is overwhelmingly high. Trends without deletions are omitted due to their high volume.}
    \label{fig:deletionpercentage}
\end{center}
\end{figure}

\descr{Deletion Percentage} High percentage of deletion is an indicator, if not the primary indicator, for the attacks as \Secref{tab:performance} shows. \Figref{fig:deletionpercentage} shows the ratio of deleted tweets to all tweets for trends with at least one deletion. The densities suggest that the percentage of deletions is overwhelmingly high for the attacked trends, even for the trends with a high number of tweets. However, there are some trends that are adopted by other users, i.e. the clients promoting themselves or their campaigns using the new trend, who post undeleted tweets and decrease this percentage. There are also some trends that are not associated with the attacks but still have a high percentage of deletions.

\begin{figure}[!htb]
\begin{center}
\includegraphics[width=\linewidth]{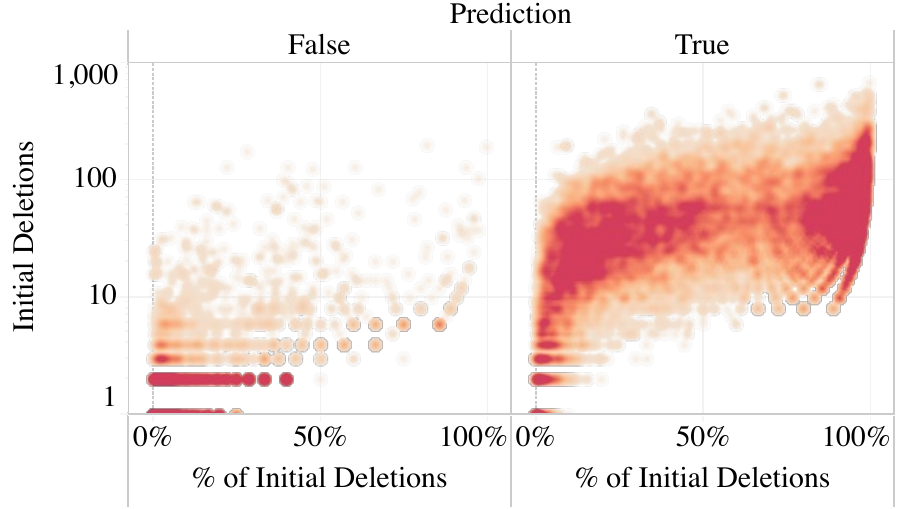}
    \caption{Initial deletions versus the percentage of tweets within the initial deletions to all tweets of the attacked trends and the other trends. Both measures are high for the attacked trends and point out the anomaly.}
    \label{fig:initialdeletions}
\end{center}
\end{figure}

\descr{Initial Deletions} In \Secref{sec:attack_analysis} we found that 72\% of all tweets associated with an astroturfed trend are deleted before they enter the trend list. However, since we do not have the exact time of trending for the whole retrospective dataset, we are not able to show this at scale. In such cases where the trends are created from scratch, we expect many attack tweets that are created and subsequently deleted before the other discussions regarding the trends take place. To approximate this phenomena, we sort the tweets and their deletions by time and we count the number of deleted single engagement tweets from the beginning until we find a tweet that is either not single engagement tweet and/or not deleted. We name those \textit{initial deletions}. We also report the ratio of the number of tweets within the initial deletions to all tweets associated with the trend. As \Figref{fig:initialdeletions} shows, the attacked trends are clearly distinct from the other trends in terms of the initial deletions. Both the initial deletions and their percentage are very low for non-attacked trends; most of them are below 3. However, for the attacked trends, not only is this number very high, but also the tweets within the initial deletions make up the majority or a sizeable minority of all tweets associated with the trend. 

\begin{figure}[ht!]
\label{fig:tweetingentropy}
\centering     
\subfigure{\label{fig:tweetingentropy_a}\includegraphics[width=\linewidth]{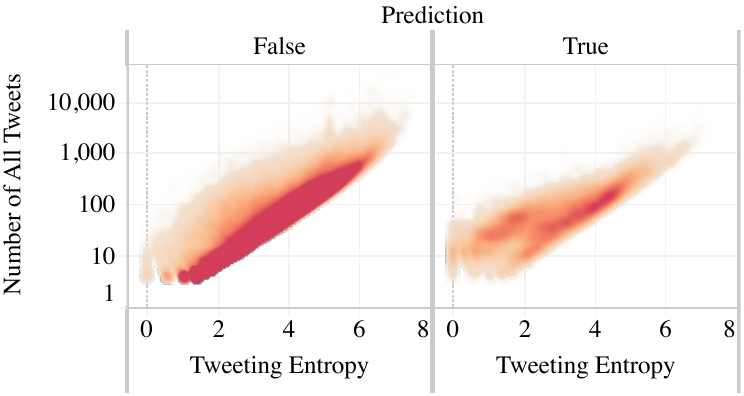}} \hfill	
\\\vspace{-15pt}
\subfigure{\label{fig:tweetingentropy_b}\includegraphics[width=\linewidth]{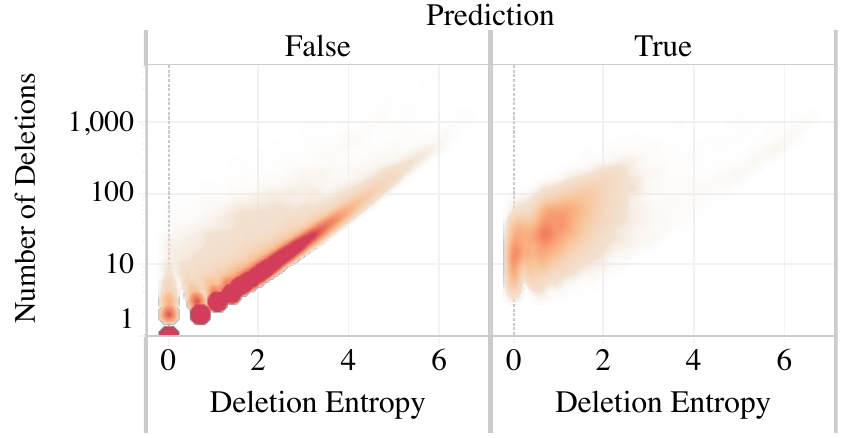}} \hfill
\caption{Entropy of the distributions of the tweets/deletions per minute. While entropy shows a linear correlation with the number of tweets/deletion for the trends that are not associated with the attacks, the attacked trends observe low entropy even with the high number of tweets/deletions.}
\end{figure}

\descr{Entropy} In \Secref{sec:turkish} we observe that the attack tweets are created in a small time window, within a minute in most cases. However it's not easy to determine the exact boundaries of the attacks due to noise. We use entropy as a proxy to show existence of such bursts of activity in a small time window. Low entropy of a distribution would indicate the distribution is concentrated in a center and therefore predictable. We compute the entropy of a trend using the number of tweets created/deleted per minute which are associated with the trend. We expect the entropy and the number of tweets to be linearly correlated since if there are more tweets/deletions, there would be more options in terms of time for them to be placed. As \Figref{fig:tweetingentropy_a} shows, this is exactly the case with the non-attacked trends. However, for the attacked trends, there is an anomalous cluster in which, despite the high number of tweets, the entropy is extremely low, i.e. there are trends which have over 50 tweets but their entropy is 0, because all tweets are created within the same minute. This behavior is observed for entropy of deletions as well, as \Figref{fig:tweetingentropy_b} shows.

\begin{figure}[!htb]
\begin{center}
\includegraphics[width=\linewidth]{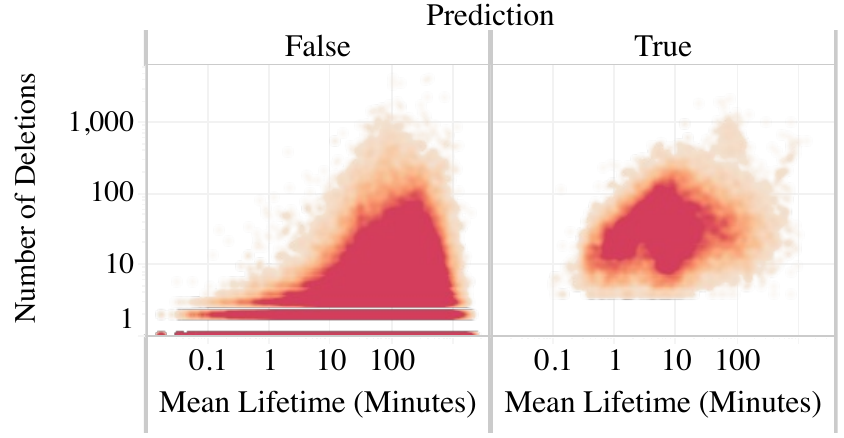}
    \caption{The mean time between deletion and creation of tweets (lifetime) per trend. Despite the high number of deletions, the mean lifetime of the attacked trends' is concentrated between 1 and 10 minutes.  }
    \label{fig:lifetimescale}
\end{center}
\end{figure}

\descr{Lifetime} We show the mean lifetime of tweets per trend. Like in the case of ground truth \Secref{sec:turkish}, the mean lifetime is very low for the attacked trends. As \Figref{fig:lifetimescale} shows, their mean lifetime is concentrated between 1 and 10 minutes even when the number of deletions is high. The non-attacked trends show a more natural behavior in which the lifetimes are not concentrated around a certain time period.

\subsection{Experimental Results of Lexicon Agnostic Classifiers}
\label{sec:experimental}

In this section, we provide the details of our experiments from \Secref{sec:turkish} on classifying the attacked trends in which we attempt to produce a more generalizable classifier.

All astroturfing attacks we inspect initiated with a burst of single engagement tweets. Additionally, the \name astroturfing attacks employ a burst of deletions after the creation of tweets which we name initial deletions. For the attacks with lexicon tweets, the single engagement tweets are lexicon tweets and classifying them helps removing noisy deleted single engagement tweets that are not part of an attack. 

To capture such behaviors, we created rules that have a high gini index using Decision Trees. These are, from most general to specific: 1) ratio of deletions to all tweets 2) number of all deletions 3) ratio of deleted non-retweets to all non-retweets 4) number of all deleted non-retweets 5) number of deleted single engagement tweets 6) ratio of deleted single engagement tweets to all single engagement tweets 7) number of initial deletions 8) number of deleted lexicon tweets 9) ratio of deleted lexicon tweets to all lexicon tweets. 

\begin{figure}[ht]
    \centering
    \includegraphics[width=\linewidth]{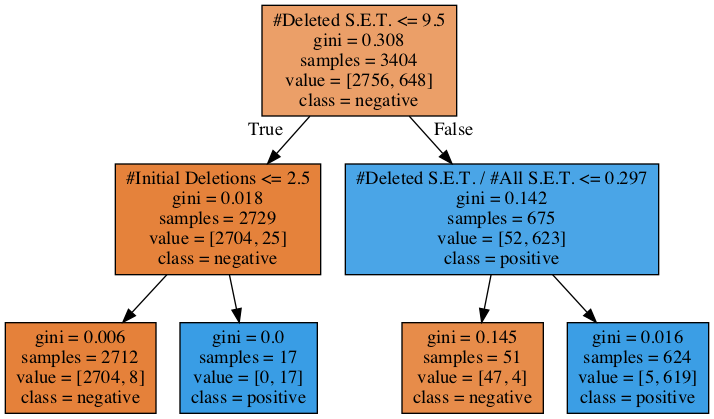}
    \caption{The lexicon agnostic classifier which achieves 97.8\% F1. 
    Number of deleted single engagement tweets and their percentage is the most important feature while initial deletions makes it possible to further classify some trends.}
    \label{fig:dtree_deletion_index}
\end{figure}

The rule 1 achieves 94.3\% 5-Fold cross validation accuracy but yields poor recall on test data (66.8\%). The content and metadata agnostic classifier employing rules 1 and 2 achieves an F-score of 78.2\% while the content agnostic classifier employing non-retweets using rules 3 and 4 achieves an F-score of 84.5\%. The lexicon agnostic decision tree using rules 5-7 ( \Figref{fig:dtree_deletion_index}) achieves 99.3\% on cross-validation accuracy and 100\% precision, 95.7\% on recall and 97.8\% on F-score. Using lexicon tweets, the decision tree on rule 8 and rule 9 performs the best with 99.7\% cross-validation score, 100\% precision, 98.9\% recall and 99.4\% F-score. Table \ref{tab:performance} shows the full results including the rules' individual performance. We conclude that the percentage of deletions among certain types of tweets is already a good indicator while the exact count of deletions is necessary to control for the unpopular trends which do not have sufficient number of tweets in the retrospective dataset, i.e. a trend with only one tweet associated with it might have a 100\% rate of deletion, but that does not mean it was attacked.

\begin{table}[ht]

\centering
\caption{The rules and the Decision Trees using them.}
\label{tab:performance}
\resizebox{\linewidth}{!}{%
\begin{tabular}{|l|l|l|l|l|l|}
\hline
Rule / Classifier                                             & 5-Fold CV       & Accuracy        & Precision      & Recall          & F1              \\ \hline
Rule 1: \#All Deletions \textgreater{}= 17                    & 92.3\%          & 84\%            & 69\%           & 51.5\%          & 59\%            \\ \hline
Rule 2: \#All Deletions / \#All Tweets \textgreater{}= 25\%   & 94.3\%          & 91.1\%          & 91.3\%         & 66.8\%          & 77\%            \\ \hline
Decision Tree: Rule 1 \& Rule 2                               & 94.9\%          & 92\%            & \textbf{100\%} & 64.2\%          & 78.2\%          \\ \hline
Rule 3: \#Deleted Non-Retweets \textgreater{}= 12             & 96\%            & 92.4\%          & 86.6\%         & 78.4\%          & 82.3\%          \\ \hline
Rule 4: \#Del. Non R.T. / \#All Non R.T. \textgreater{}= 0.34 & 96.1\%          & 93.1\%          & 84.7\%         & 84.7\%          & 84.7\%          \\ \hline
Decision Tree: Rule 3 \& Rule 4                               & 97.4\%          & 94\%            & \textbf{100\%} & 73.1\%          & 84.5\%          \\ \hline
Rule 5: \#Deleted S.E.T. \textgreater{}= 10                   & 97.3\%          & 96\%            & 93.8\%         & 87.8\%          & 90.7\%          \\ \hline
Rule 6: \#Deleted S.E.T / \#All S.E.T \textgreater{}= 50\%    & 97.3\%          & 95.5\%          & 89.5\%         & 90.5\%          & 90\%            \\ \hline
Rule 7: \#Initial Deletions \textgreater{}= 4                 & 94.8\%          & 92.2\%          & 98.4\%         & 66.3\%          & 79.2\%          \\ \hline
\textbf{Decision Tree: Rule 5 \& Rule 6 \& Rule 7}            & 99.3\%          & 99\%            & \textbf{100\%} & 95.7\%          & 97.8\%          \\ \hline
Rule 8: \#Deleted Lex. Tw. \textgreater{}= 4                  & \textbf{99.7\%} & 99.5\%          & 98.4\%         & \textbf{99.4\%}          & 98.9\%          \\ \hline
Rule 9: \#Del. Lex. Tw. / \#All Lex. Tw. \textgreater{}= 68\% & 98.3\%          & 97.1\%          & 91\%           & 96.8\% & 93.8\%          \\ \hline
\textbf{Decision Tree: Rule 8 \& Rule 9}                      & \textbf{99.7\%} & \textbf{99.7\%} & \textbf{100\%} & 98.9\%          & \textbf{99.4\%} \\ \hline
\end{tabular}%
}
\end{table}

\subsection{Other Countries}
\label{sec:other_countries}
The Decision Tree working on the rules 5-7 from the previous section is the best lexicon agnostic classifier to classify \name astroturfing attacks when a different kind of generated content is employed instead of lexicon tweets. We used this classifier for classification of trends from other countries in 2019. The left-hand side of the Decision Tree in \Figref{fig:dtree_deletion_index} did not yield any positives, meaning that there is no hashtag that starts with 3 single engagement tweets which are deleted within the same day. The right-hand side of the tree yields 120 hashtags that trended over 1,124 days. Some of these hashtags are "\#NewProfilePic" in various languages. Only 38 of these hashtags trended. Three of them are about popular TV shows but with a typo: \#GamesOfThrones, \#Loveisalnd, \#loveisand. We believe that these are false positives in which many users attempted to tweet about this TV shows at the same time with a typo and deleted their tweets but could not help pushing the keyword with typo to trends. The others are in foreign languages (mostly Arabic) and appear to be common spam rather than attacks. 

Additionally, to show that only Turkish trends suffer from \name astroturfing attacks, we came up with the following methodology:

\begin{figure}[ht]
    \centering
      \includegraphics[width=\columnwidth]   {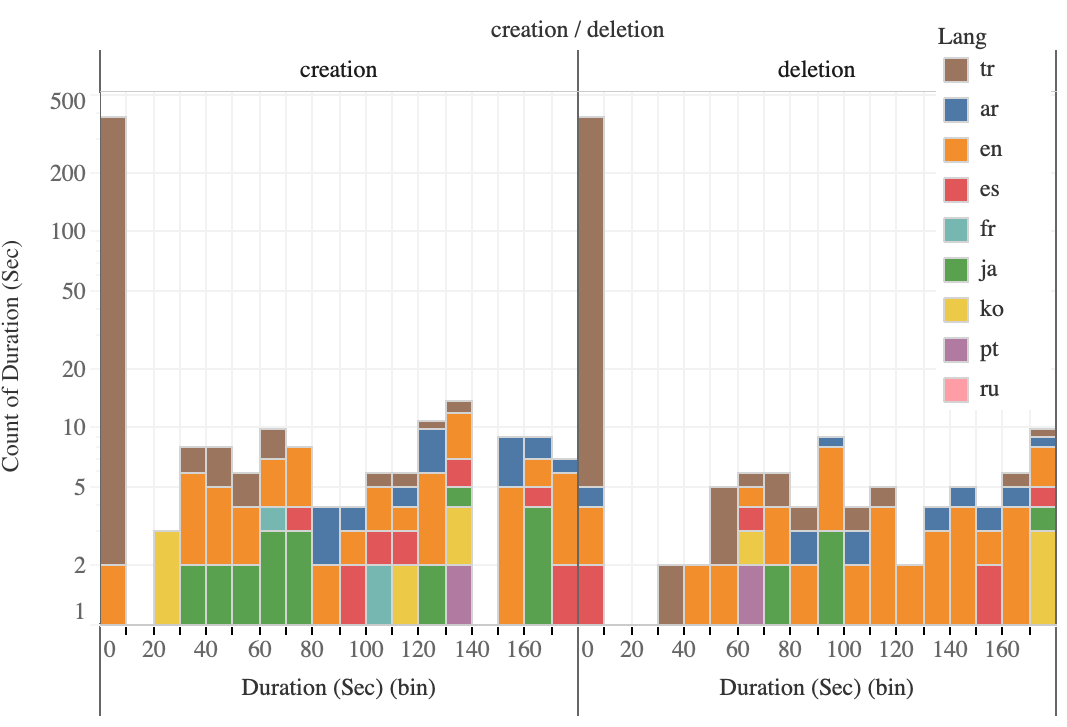}
    \caption{Duration of the attack (in seconds) is the difference between the creation times (left) / the deletion times (right) of consecutive tweets associated with a trend. The median time of the duration shows many Turkish trends with this anomaly, while only a small number of non-Turkish trends have such behavior.}
    \label{fig:duration_other}
\end{figure}

\begin{figure}[ht]
    \centering
    \includegraphics[width=0.8\columnwidth]  {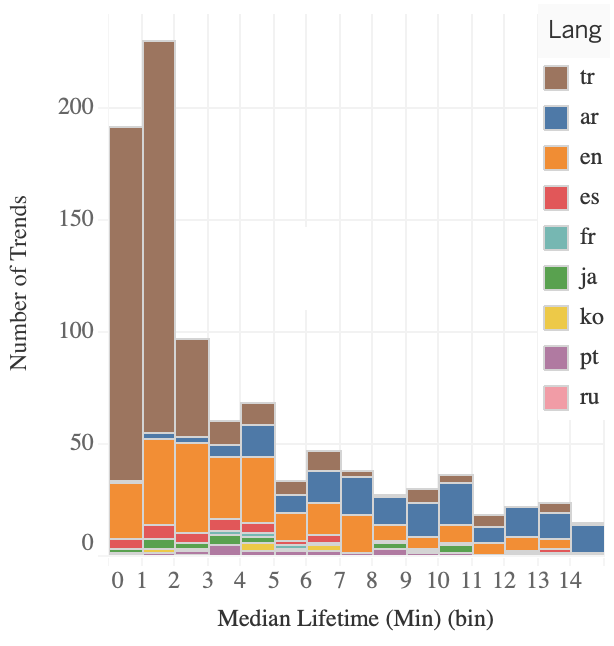}
    \caption{Median Lifetime (Min) is the time that the tweets stay on the platform before got deleted (if they are deleted). Median lifetime of tweets associated with Turkish trends are generally very low while such anomaly is not observed prevalently in non-Turkish trends.}
    \label{fig:lifetime_other}
\end{figure}

We search for the pattern that we observed in Turkey: 1) many tweets created and deleted in a very small timeframe 2) many tweets with a short lifetime, and 3) many tweets are deleted. To measure 1), we compute the median difference between the creations/deletion times of each pair of adjacent tweets for each trend. For 2), we measure the median lifetime of each tweet associated with each trend. Lastly, we record whether or not a majority of the tweets associated with a trend were eventually deleted in the same day. We only consider S.E.Ts. We show the results on the world trends between 1st June 2019 - 31st July 2019 by the most frequent language of the tweets using them. 

\begin{figure}[hb]
    \centering
    \includegraphics[width=\columnwidth]  {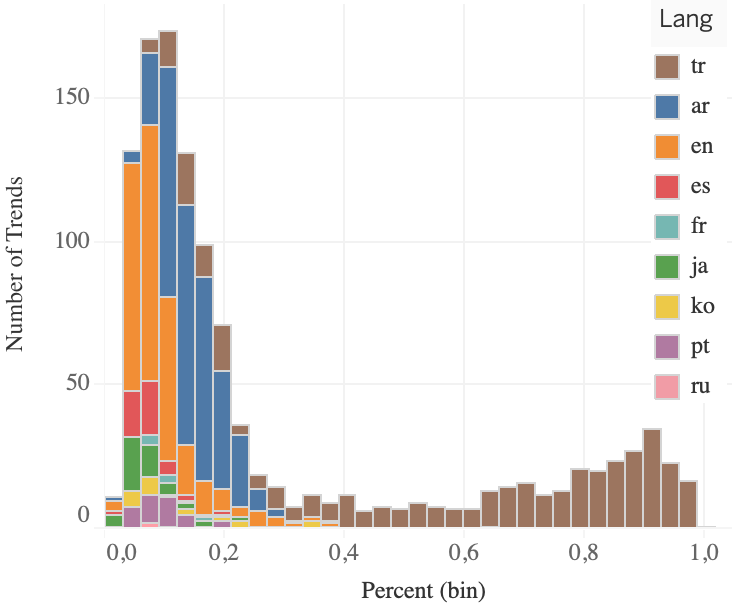}
    \caption{Percentage of tweets associated with a trend that are deleted. Only Turkish trends have at least 40\% of tweets that are deleted with only one exception.}
    \label{fig:percentage_other}
\end{figure}

\Figref{fig:duration_other} shows that a large portion of Turkish trends have tweets that were created in a small window and then deleted in a small window. The median time between consecutive creation and deletions are below 10 seconds as we also found out in ~\Secref{sec:turkish}. This is rarely the case with trends in other languages. \Figref{fig:lifetime_other} shows that many Turkish trends have very small lifetime. Again, we do not see this behaviour on trends in other languages. Finally, \Figref{fig:percentage_other} shows that only Turkish trends have tweets in which more than 40\% deleted with only one exception. There are no single trend that have all the anomalous patterns the Turkish trends have.

\end{appendix}
\end{document}